\documentclass[a4paper,onecolumn,nofootinbib]{revtex4}

\usepackage[top=2.8cm, bottom=2.8cm, left=2.4cm, right=2.4cm]{geometry}
\usepackage[utf8]{inputenc}  
\usepackage[T1]{fontenc}     
\usepackage{amsmath,amssymb,lmodern} 
\usepackage{graphicx}
\usepackage{diagbox}
\usepackage{hyperref}
\usepackage{color}
\usepackage{lipsum}
\usepackage{bbm}
 \usepackage{young}

\usepackage{xcolor}
\definecolor{blue}{rgb}{0.19,0.64,0.54}
\definecolor{reddish}{rgb}{0.65, 0.2, 0.2}
\definecolor{red}{rgb}{0.7,0.3,0.3}
\definecolor{darkgreen}{rgb}{0.2,0.7,0.3}
\definecolor{darkblue}{rgb}{0.3,0.40,0.48}
\definecolor{gray}{rgb}{.8,.8,.8}

\hypersetup{,
    unicode=false,          
    pdftoolbar=true,        
    pdfmenubar=true,        
    pdffitwindow=false,     
    pdfstartview={FitH},    
    pdftitle={Generalized Proca SU(2)},    
    pdfauthor={E. Allys,},     
    pdfsubject={Subject},   
    pdfcreator={Creator},   
    pdfproducer={Producer}, 
    pdfkeywords={keyword1} {key2} {key3}, 
    pdfnewwindow=true,      
    colorlinks=true,       
    linkcolor=red,          
    citecolor=darkblue,        
    filecolor=magenta,      
    urlcolor=darkblue,           
    linktocpage=true
}

\renewcommand{\a}{{\color{red}a\color{black}}}
\renewcommand{\b}{{\color{red}b\color{black}}}
\renewcommand{\c}{{\color{red}c\color{black}}}
\renewcommand{\d}{{\color{red}d\color{black}}}
\newcommand{\e}{{\color{red}e\color{black}}}
\newcommand{\f}{{\color{red}f\color{black}}}
\newcommand{\g}{{\color{red}g\color{black}}}

\newcommand{\al}{{\color{red}\alpha\color{black}}}
\newcommand{\be}{{\color{red}\beta\color{black}}}
\newcommand{\ga}{{\color{red}\gamma\color{black}}}
\newcommand{\au}{{\color{red}a_1\color{black}}}
\newcommand{\am}{{\color{red}a_m\color{black}}}
\newcommand{\aN}{{\color{red}a_N\color{black}}}

\begin{document}
\title{New terms for scalar multi-Galileon models and application to SO($N$) and SU($N$) group representations}

\author{Erwan Allys}
\email{allys@iap.fr}
\affiliation{$\mathcal{G}\mathbb{R}\epsilon\mathbb{C}\mathcal{O}$, Institut d’Astrophysique de Paris, UMR 7095, 98 bis bd Arago, 75014 Paris, France \\ UPMC Universit\'e Paris 6 et CNRS,
  Sorbonne Universit\'es}
\date{\today}

\begin{abstract}
We investigate a new class of scalar multi-Galileon models, which is not included in the commonly admitted general formulation of generalized multi-Galileons. The Lagrangians of this class of models, some of them having already been introduced in previous works, are specific to multi-Galileon theories, and vanish in the single Galileon case. We examine them in detail, discussing in particular some hidden symmetry properties which can be made explicit by adding total derivatives to these Lagrangians. These properties allow us to describe the possible dynamics for these new Lagrangians in the case of multi-Galileons in the fundamental representation of a SO($N$) and SU($N$) global symmetry group, as well as in the adjoint representation of a SU($N$) global symmetry group. We perform in parallel an exhaustive examination of some of these models, finding a complete agreement with the dynamics obtained using the symmetry properties. Finally, we conclude by discussing what could be the most general multi-Galileon theory, as well as the link between scalar and vector multi-Galileon models.
\end{abstract}

\maketitle
\section{Introduction}

In recent years, the attempts to investigate in a systematic way the different classes of modified gravity theories have been very successful. Galileon theories, describing models involving one scalar field coupling to general relativity, have been extensively studied \cite{Nicolis:2008in,Deffayet:2009wt,Deffayet:2009mn,Deffayet:2011gz,deRham:2011by,Joyce:2014kja}. Their most general extension has been especially proven in Ref.~\cite{Deffayet:2011gz}. This approach to modify gravity has found multiple applications in cosmology, for example on the subject of dark energy \cite{Chow:2009fm,Silva:2009km,Deffayet:2010qz,Kobayashi:2010wa,Gannouji:2010au,Tsujikawa:2010zza,DeFelice:2010pv,Ali:2010gr,DeFelice:2010nf,Mota:2010bs,Nesseris:2010pc,Easson:2011zy,Gleyzes:2013ooa,Gabadadze:2016llq,Salvatelli:2016mgy,Shahalam:2016kkg,Minamitsuji:2016qyc,Saridakis:2016ahq,Biswas:2016bwq} or inflation \cite{Creminelli:2010ba,Kobayashi:2010cm,Mizuno:2010ag,Burrage:2010cu,Creminelli:2010qf,Kamada:2010qe,Libanov:2016kfc,Banerjee:2016hom,Hirano:2016gmv,Brandenberger:2016vhg,Nishi:2016wty}. It has even been shown that some Galileon models are as compatible with cosmological data as the $\Lambda$CDM model \cite{Neveu:2016gxp}.

Several attempts have been made to investigate theories going beyond the standard single scalar Galileon theory. For example, the possibility to build vector Galileon models with vector fields propagating three degrees of freedom have been investigated \cite{Deffayet:2010zh,Heisenberg:2014rta,Tasinato:2014eka,Allys:2015sht,Jimenez:2016isa,Allys:2016jaq}, as well as their first cosmological applications \cite{Tasinato:2014eka,Tasinato:2014mia,Hull:2014bga,DeFelice:2016cri,DeFelice:2016yws,DeFelice:2016uil,Heisenberg:2016wtr}. The possibility to have several scalar fields has also been investigated \cite{Padilla:2010de,Padilla:2010tj,Padilla:2010ir,Zhou:2010di,Hinterbichler:2010xn,Andrews:2010km,Padilla:2012dx,Sivanesan:2013tba,Garcia-Saenz:2013gya,Charmousis:2014zaa,Saridakis:2016mjd}. Such models are called multi-Galileon ones, and can be considered for arbitrary internal indices, as well as for multi-Galileons in given group representations. A formulation of what would be the most general multi-Galileon theory has been especially discussed in Refs.~\cite{Padilla:2012dx,Sivanesan:2013tba}. However, it has been shown that some models are not included in this previously derived general action, for example the multifield Dirac-Born-Infeld Galileons~\cite{Kobayashi:2013ina}. Additional terms have also been derived e.g. in Ref.~\cite{Deffayet:2010zh} for arbitrary $p$-forms or Ref.~\cite{Ohashi:2015fma} for bi-Galileon theory that do not enter this general class of models. All these extra terms are included in the present construction.

In this paper, we discuss a class of terms satisfying the standard multi-Galileon hypotheses, but which are not included in the general action of Refs.~\cite{Padilla:2012dx,Sivanesan:2013tba}. These Lagrangians, which we call extended multi-Galileon ones, are specific to multi-Galileon models, and identically vanish in the single Galileon case. In the first part, we introduce these new Lagrangians, and examine their different properties, including hidden symmetry properties, i.e. which can be made explicit by adding conserved currents to the Lagrangians. Then, in Secs.~\ref{PartFundRep} and~\ref{PartAdjRep}, we use the previous properties to investigate all the possible dynamics for extended multi-Galileon Lagrangians in the fundamental representation of a SO($N$) or SU($N$) global symmetry group, and in the adjoint representation of a SU($N$) global symmetry group. A similar work has previously been done in~\cite{Padilla:2010ir} for multi-Galileon Lagrangians with equations of motion of order two only. For some of these models, we also perform in parallel an exhaustive examination of all the possible Lagrangians. The results of these systematic investigations, mostly given in Appendices~\ref{AppEpsSON} and~\ref{AppPartSUN}, are in complete agreement with the dynamics obtained using the complete symmetry properties of the Lagrangians, which strengthens our examination. This investigation also allows us to examine the internal properties of the model, e.g. the link between the possible alternative Lagrangian formulations. We conclude the paper in Sec.~\ref{PartConclusion}, in particular by discussing what could be the most general multi-Galileon theory, as well as the link between scalar and vector multi-Galileon models.

\section{Extended multi-Galileon Theory}
\subsection{Presentation}
\label{IntroMultiGal}

We discuss in this paper possible Lagrangian terms for multi-Galileon theories in flat spacetime (which we assume is four-dimensional). These terms are built from multi-Galileon fields only (which we also call simply multi-Galileons), i.e. scalar fields with internal indices $\pi^\a$. These fields can lie in given group representations, or only be parts of generic nonlinear sigma models. The multi-Galileon theories satisfy the following conditions
\begin{itemize}
\item[i)] \textsl{The Lagrangians contain up to second-order derivatives of the multi-Galileons.}
\item[ii)] \textsl{The Lagrangians are polynomial in the second-order derivatives of the multi-Galileons.}
\item[iii)] \textsl{The field equations contain up to second-order derivatives of the multi-Galileon fields.}
\end{itemize}
The third condition is necessary in order for the theory not to include the Ostrogradski instability~\cite{Woodard:2006nt,Woodard:2015zca}. See also~\cite{Motohashi:2014opa} for a discussion of the instability coming from the third-order derivatives in the equations of motion, especially in the case of multifield theories. It has been proven in Ref.~\cite{Sivanesan:2013tba} that the conditions \textbf{i} and \textbf{iii} imply the condition \textbf{ii}. We however leave it as a hypothesis here.

The starting point for the study of such theories has been the single Galileon case, which has already been examined in detail~\cite{Nicolis:2008in,Deffayet:2009wt,Deffayet:2009mn,Deffayet:2011gz,deRham:2011by}. The generalized Galileon theory has been proven to be the most general one with the hypotheses given below, when considering the case of a unique scalar field~\cite{Deffayet:2011gz}. Its construction begins from the most general theory giving  only second-order equations of motion, i.e. Lagrangians of the form
\begin{equation}
\displaystyle{\mathcal{L}^{\text{Gal}}_0 = \delta^{\mu_1 \cdots \mu_m}_{\nu_1\cdots\nu_m} \pi
\partial_{\mu_1}\partial^{\nu_1} \pi \cdots \partial_{\mu_m}\partial^{\nu_m} \pi,}
\label{EqLagGalIni0}
\end{equation}
or equivalently, up to a total derivative,
\begin{equation}
\displaystyle{\mathcal{L}^{\text{Gal}}_1 = \delta^{\mu_1 \cdots \mu_m}_{\nu_1\cdots\nu_m} \partial_{\mu_1} \pi
\partial^{\nu_1} \pi \partial_{\mu_2}\partial^{\nu_2}  \pi \cdots \partial_{\mu_m}\partial^{\nu_m} \pi,}
\label{EqLagGalIni1}
\end{equation}
with $m$ taking values between $1$ and $4$, and where 
\begin{equation}
\displaystyle{\delta^{j_1 \cdots j_n}_{i_1 \cdots
i_n} =  n! \delta^{j_1
\cdots j_n}_{[i_1 \cdots i_n]} = \frac{1}{(D-n)!}\epsilon^{i_1\cdots i_n \sigma_1 \cdots \sigma_{D-n}}
\epsilon_{j_1\cdots j_n \sigma_1 \cdots \sigma_{D-n}} =   \delta^{j_1}_{i_1} \cdots \delta^{j_n}_{i_n} \pm \cdots,}
\label{EqDeltaMult}
\end{equation}
for $n$ running from 1 to 4 (in a four-dimensional spacetime). Then, multiplying those terms by an arbitrary function of $\pi$ and its first derivative still gives second-order equations of motion. This is due to the fact that all the third-order derivatives in the equations of motion coming from the variation of the first-order derivatives in the arbitrary function and the second-order derivatives in the initial Galileon Lagrangian will cancel each other out. Indeed, both terms produce the same third-order derivative contribution, but from a different number of integrations by parts, one and two respectively, granting them an opposite sign (see Ref.~\cite{Deffayet:2011gz} for a detailed discussion of this property).

Multi-Galileon Lagrangians have been obtained in the same way. The Lagrangians giving only second-order equations of motion were first examined in Refs.~\cite{Deffayet:2010zh,Padilla:2010de,Padilla:2010tj}, and consist in adding internal indices to the Galileon Lagrangians of Eqs.~\eqref{EqLagGalIni0} or \eqref{EqLagGalIni1}. One could then consider the possibility to multiply these Lagrangians by an arbitrary function of the multi-Galileons and their first-order derivatives. However, the property of cancellation of third-order derivative terms discussed in the single Galileon case is not valid anymore, since the internal indices can break the symmetry between the contributions to the equations of motion coming from one or two integrations per part, respectively. Thus, the way to overcome this difficulty is to ensure the necessary symmetry between these pairs of terms~\cite{Sivanesan:2013tba,Padilla:2012dx}. Following this procedure, the related Lagrangians are of the form 
\begin{equation}
\displaystyle{\mathcal{L}^{\text{multiGal}} = A^{\au\cdots \am}\left(X_{\a\b},\pi_\c \right) 
\delta^{\mu_1 \cdots \mu_m}_{\nu_1\cdots\nu_m} 
\partial_{\mu_1}\partial^{\nu_1} \pi_{\au} \cdots \partial_{\mu_m}\partial^{\nu_m} \pi_{\am},}
\label{EqMultiGalVish}
\end{equation}
with $X_{\a\b}= (1/2) \partial_\rho \pi_\a \partial^\rho \pi_\b$, and with the property that $\partial A^{\au\cdots \am} /\partial X_{\a\b}$ is symmetric in all of its indices $(\au, \cdots,\am,\a,\b)$. The theory spanned by these Lagrangian was determined to be the most general one in Refs.~\cite{Sivanesan:2013tba,Padilla:2012dx}.

However, some authors found that other terms verifying the same hypotheses \textbf{i}-\textbf{iii} are not included in these Lagrangians~\cite{Kobayashi:2013ina,Ohashi:2015fma}. One could then ask if the construction done by multiplying the Lagrangians giving second-order-only equations of motion by such an arbitrary function, while still verifying the above conditions \textbf{i}-\textbf{iii} produce all the possible terms. Indeed, including this arbitrary function which satisfies the symmetry condition introduced in Refs.~\cite{Sivanesan:2013tba,Padilla:2012dx} is a sufficient condition to include extra first-order derivatives in the action, but not a necessary one. Another way to incorporate additional first-order derivatives into such a Lagrangian while keeping second-order equations of motion, if the Lagrangian does not already contain more than two second-order derivatives, is to use the antisymmetry of the term given in Eq.~\eqref{EqDeltaMult}. We can indeed write the following Lagrangians, introduced especially in Ref.~\cite{Deffayet:2010zh}:
\begin{equation}
\displaystyle{\begin{array}{l}
\mathcal{L}^{\text{ext}}_{_I} =  \delta^{\mu_1 \mu_2 \mu_3}_{\nu_1\nu_2\nu_3} \partial_{\mu_1}\pi^\a  
\partial_{\mu_2}\pi^\b\partial^{\nu_1}\pi^\c\partial^{\nu_2}\pi^\d \partial_{\mu_3}\partial^{\nu_3}\pi^\e,\\
\mathcal{L}^{\text{ext}}_{_{II}} =  \delta^{\mu_1 \mu_2 \mu_3 \mu_4}_{\nu_1\nu_2\nu_3\nu_4} \partial_{\mu_1}\pi^\a  \partial_{\mu_2}\pi^\b\partial^{\nu_1}\pi^\c\partial^{\nu_2}\pi^\d \partial_{\mu_3}\partial^{\nu_3}\pi^\e\partial_{\mu_4}\partial^{\nu_4}\pi^\f,\\
\mathcal{L}^{\text{ext}}_{_{III}} =  \delta^{\mu_1 \mu_2 \mu_3 \mu_4}_{\nu_1\nu_2\nu_3\nu_4} \partial_{\mu_1}\pi^\a  \partial_{\mu_2}\pi^\b\partial_{\mu_3}\pi^\c\partial^{\nu_1}\pi^\d\partial^{\nu_2}\pi^\e\partial^{\nu_3}\pi^\f \partial_{\mu_4}\partial^{\nu_4}\pi^\g,
\end{array}}
\label{EqExtGal}
\end{equation}
where we leave the internal indices totally free for the moment. We call these particular Lagrangians extended multi-Galileon ones. 

Due to the contraction of all derivative terms with a prefactor of the form $\delta^{j_1 \cdots j_n}_{i_1 \cdots i_n} $, it is straightforward to see that these Lagrangians give second-order equations of motion. In addition, their particular structure, implying antisymmetric properties between the internal indices, makes them different from the Lagrangians of Eq.~\eqref{EqMultiGalVish}. To see this, one can consider the case of $\mathcal{L}^{\text{ext}}_{_I}$. Terms of the same order from Lagrangians of the form of Eq.~\eqref{EqMultiGalVish} are $\mathcal{L}_1 = \partial_\mu\pi^\a \partial^\mu\pi^\b \partial_\nu \pi^\c \partial^\nu \pi^\d \partial_\rho\partial^\rho \pi^\e$ and $\mathcal{L}_2 = \pi ^\a\partial_\mu\pi^\b \partial^\mu\pi^\c   \delta^{\mu_1 \mu_2 }_{\nu_1\nu_2} \partial_{\mu_3}\partial^{\nu_3}\pi^\d\partial_{\mu_4}\partial^{\nu_4}\pi^\e$. They have to be symmetric in all the internal indices for the first one, and all the internal indices but $\a$ in the second, which means that an antisymmetry under the exchanges of two pairs of indices is not possible. In addition, they do not vanish in the single Galileon limit, whereas $\mathcal{L}^{\text{ext}}_{_I}$ vanishes in this limit. It means that they can be obtained by adding internal indices to single Galileon models, which is not the case for $\mathcal{L}^{\text{ext}}_{_I}$, making this last Lagrangian a purely multi-Galileon one.

These extended Lagrangians can be considered as a basis, in addition to the Lagrangians giving second-order-only equations of motion, to build the generalized multi-Galileon models. Indeed, it is still possible to include an arbitrary function of $\pi^\a$ and its first derivative in front of them, as long as one ensures that the third-orders derivatives appearing in the equations of motion cancel each other out thanks to additional symmetry properties of the internal indices. Note that this kind of generalized multi-Galileon Lagrangians was already partially written in the literature. In the following sections, we will focus on the extended multi-Galileon Lagrangians only.

\subsection{General properties}
\label{PartPropGen}

Let us first consider the symmetry properties of these extended Lagrangians. The properties of $\delta^{j_1 \cdots j_n}_{i_1 \cdots i_n} $ impose that the Lagrangians are completely antisymmetric by exchange of the fields with $\partial_{\mu_i}$ derivatives only, as well as by exchange of the fields with $\partial^{\nu_i}$ derivatives only. They are also symmetric by exchange of the groups of fields with $\partial_{\mu_i}$ derivatives only and $\partial^{\nu_i}$ derivatives only, since the $\mu_i$ and $\nu_i$ can be exchanged without modifying the Lagrangian. Finally, $\mathcal{L}^{\text{ext}}_{_{II}}$ is symmetric by exchange of the fields with second-order derivatives. To summarize, and taking the example of 
\begin{equation}
\displaystyle{\mathcal{L}^{\text{ext}}_{_{II}} =  \delta^{\mu_1 \mu_2 \mu_3 \mu_4}_{\nu_1\nu_2\nu_3\nu_4} \partial_{\mu_1}\pi^\a  \partial_{\mu_2}\pi^\b\partial^{\nu_1}\pi^\c\partial^{\nu_2}\pi^\d \partial_{\mu_3}\partial^{\nu_3}\pi^\e\partial_{\mu_4}\partial^{\nu_4}\pi^\f,}
\end{equation}
this Lagrangian is 
\begin{itemize}
\item[i)] \textsl{Antisymmetric under the exchanges $\a\leftrightarrow \b$ or $\c\leftrightarrow\d$},
\vspace{-0.3cm}
\item[ii)] \textsl{Symmetric under the exchange $\e\leftrightarrow\f$},
\vspace{-0.3cm}
\item[iii)] \textsl{Symmetric under exchange $(\a,\b)\leftrightarrow (\c,\d)$}.
\end{itemize}
These symmetries are very restrictive, particularly for models where there are only a few internal states. For example, in the case of a bi-Galileon model, and for $\mathcal{L}^{\text{ext}}_{_I}$, only two independent terms are possible out of the 32 initial configurations, i.e. $\mathcal{L}_1=  \delta^{\mu_1 \mu_2 \mu_3}_{\nu_1\nu_2\nu_3} \partial_{\mu_1}\pi^{\color{red}1\color{black}}
\partial_{\mu_2}\pi^{\color{red}2\color{black}}\partial^{\nu_1}\pi^{\color{red}1\color{black}}\partial^{\nu_2}\pi^{\color{red}2\color{black}} \partial_{\mu_3}\partial^{\nu_3}\pi^{\color{red}1\color{black}}$ and $\mathcal{L}_2=  \delta^{\mu_1 \mu_2 \mu_3}_{\nu_1\nu_2\nu_3} \partial_{\mu_1}\pi^{\color{red}1\color{black}}
\partial_{\mu_2}\pi^{\color{red}2\color{black}}\partial^{\nu_1}\pi^{\color{red}1\color{black}}\partial^{\nu_2}\pi^{\color{red}2\color{black}} \partial_{\mu_3}\partial^{\nu_3}\pi^{\color{red}2\color{black}}$.

Another symmetry of this Lagrangian can be seen indirectly, with the following property:
\begin{itemize}
\item[\textbf{a)}] \textsl{If there is an antisymmetry between the internal indices of fields with first and second-order derivatives in a Lagrangian $\mathcal{L}^{\text{ext}}_n$, this Lagrangian is a total derivative. In the case of $\mathcal{L}^{\text{ext}}_{_I}$, it would be for example an antisymmetry between fields with indices $\a$ and $\e$, or $\c$ and $\e$.}
\end{itemize}
Let us prove it in the case of  $\mathcal{L}^{\text{ext}}_{_I}$. For this purpose, we can consider the contraction of the internal indices with a function which contains no fields. This function can be arbitrary in a general sigma model, and has to be built from primitive invariant tensors in the case of multi-Galileons in a given group representation. This function has to verify the symmetry properties detailed previously. So, we can write 
\begin{equation}
\label{EqLExt1Cont}
\displaystyle{\mathcal{L}^{\text{ext}}_{I,A} = \delta^{\mu_1 \mu_2 \mu_3}_{\nu_1\nu_2\nu_3} \partial_{\mu_1}\pi^\a  
\partial_{\mu_2}\pi^\b\partial^{\nu_1}\pi^\c\partial^{\nu_2}\pi^\d \partial_{\mu_3}\partial^{\nu_3}\pi^\e A_{[\a\b][\c\d]\e},}
\end{equation}
where the square brackets mean antisymmetrization. Suppose that in addition to these symmetry properties, there is an antisymmetry of $A$ under the exchange $\c\leftrightarrow\e$, without loss of generality. In this case, $A$ is antisymmetric on the group of indices $(\c,\d,\e)$. Indeed, the antisymmetry under the exchange $\b\leftrightarrow\c$ is forced, since the symmetric configuration always vanishes. Then, the only possibility for a tensor that is antisymmetric under the exchanges $\c\leftrightarrow\d$ and $\c\leftrightarrow\e$ is to be completely antisymmetric on these three indices. It is then possible to build the following current
\begin{equation}
\displaystyle{J^{\mu_1} = \delta^{\mu_1 \mu_2 \mu_3}_{\nu_1\nu_2\nu_3} \pi^\a  
\partial_{\mu_2}\pi^\b\partial^{\nu_1}\pi^\c\partial^{\nu_2}\pi^\d \partial_{\mu_3}\partial^{\nu_3}\pi^\e A_{[\a\b][\c\d]\e}},
\end{equation}
which is not vanishing if the Lagrangian is not vanishing. 
Using the symmetry properties of $A$, one can compute that
\begin{equation}
\displaystyle{\partial_{\mu_1} J^{\mu_1} = \mathcal{L}^{\text{ext}}_{1,a} +2 \delta^{\mu_1 \mu_2 \mu_3}_{\nu_1\nu_2\nu_3}  \pi^\a  \partial_{\mu_2}\pi^\b\partial_{\mu_1}\partial^{\nu_1}\pi^\c\partial^{\nu_2}\pi^\d \partial_{\mu_3}\partial^{\nu_3}\pi^\e A_{[\a\b][\c\d]\e}.}
\end{equation}
However, the second term vanishes, since $A$ is antisymmetric by exchange of the indices $\c\leftrightarrow\e$, which proves that the $\mathcal{L}^{\text{ext}}_{1,a}$ is a total derivative. Another way to prove this property is by introducing the current 
\begin{equation}
\displaystyle{J^{\mu_3} = \delta^{\mu_1 \mu_2 \mu_3}_{\nu_1\nu_2\nu_3} \partial_{\mu_1} \pi^\a  
\partial_{\mu_2}\pi^\b\partial^{\nu_1}\pi^\c\partial^{\nu_2}\pi^\d \partial^{\nu_3}\pi^\e A_{[\a\b][\c\d]\e}},
\end{equation}
whose divergence gives 3 times the expected Lagrangian. This current clearly shows that it is the contribution totally antisymmetric here in $(\c,\d,\e)$ which is a total derivative, since this is the only one which gives a nonvanishing $J^{\mu_3}$. A similar calculation can be done for the other extended multi-Galileon Lagrangians.

This allows us to prove another symmetry property, which will turn out to be very useful in the following:
\begin{itemize}
\item[\textbf{b)}] \textsl{The extended multi-Galileon Lagrangians with a nontrivial dynamics can be written up to a total derivative in a form that has a complete symmetry under the exchange of the internal indices of one field with a single $\partial_{\mu_j}$ derivative, one field with a single $\partial^{\nu_j}$ derivative, and all the fields with a $\partial_{\mu_k} \partial^{\nu_k}$ derivative, and no other fields. Without loss of generality, we can assume this symmetry between the fields with $\partial_{\mu_1}$, $\partial^{\nu_1}$, $\partial_{\mu_3} \partial^{\nu_3}$ and $\partial_{\mu_4} \partial^{\nu_4}$ (if it exists) derivatives in the Lagrangians of Eq.~\eqref{EqExtGal}. In this form, the other indices of fields with one derivatives can be symmetrized by pairs, e.g. the fields with $\partial_{\mu_2}$ and $\partial^{\nu_2}$ derivatives, and so on.}
\end{itemize}
Two independent proofs of this property are given in Appendix~\ref{AppProofB}.

An important remark is that property~\textbf{b} does not claim that the Lagrangians which have a nontrivial dynamics possess these particular symmetry properties, but that they can be written as Lagrangians with these symmetry properties plus antisymmetric total derivatives. In other words, the complete symmetry of a Lagrangian can be hidden by total derivatives. It is linked to the fact that starting from such a symmetric Lagrangian, it is always possible to add conserved currents which do not possess these symmetries. On the other hand, this property shows that it is sufficient to look for Lagrangians which have these complete symmetry properties.

Finally, a last property of these extended multi-Galileon Lagrangians can be proven:
\begin{itemize}
\item[\textbf{c)}] \textsl{The equations of motion for the extended multi-Galileon Lagrangians are total derivatives.}
\end{itemize}
It can be shown from the fact that these Lagrangians contain only derivatives of the scalar fields. The equations of motion have thus the following form,
\begin{equation}
\displaystyle{0= \left[\partial_\mu \partial_\nu \left(\frac{\partial}{\partial (\partial_\nu \partial_\mu \pi_\al)} \right)- \partial_\mu \left(\frac{\partial}{\partial (\partial_\mu \pi_\al)}\right) \right] \mathcal{L}^{\text{ext}}_n=
\partial_\mu J^{\mu\al}}
\end{equation}
This last property is useful to verify the consistency of calculations. Indeed, the generalized Galileon actions obtained by multiplying the extended multi-Galileon Lagrangians by functions of $\pi^\a$ and its first derivative do not verify this property anymore.

\subsection{Extended multi-Galileon in group representations}
\label{PartPropGroup}

In the following, to reduce the possible contractions between the internal indices, we will work with multi-Galileons lying in a given group representation. We suppose that this group transformation describes a global symmetry of the model, from an effective field theory point of view. Thus, we assume that the Lagrangians behave as singlets under the action of this symmetry group. The representations we will take as examples will be the fundamental representation of a SO($N$) or SU($N$) symmetry group, and the adjoint representation of a SU($N$) symmetry group. A similar work has previously been done in Ref.~\cite{Padilla:2010ir} for multi-Galileon Lagrangians with equations of motion of order two only.

In these representations, it is possible to write the Lagrangians as a contraction between the terms given in Eq.~\eqref{EqExtGal} and a prefactor term which is built only from primitive invariant tensors of the multi-Galileon representation, simply called primitive invariants in the following. The primitive invariants of a given group representation are the set of invariant tensors (under the action of the group) with the minimal possible numbers of indices and from which all invariant tensors can be built, using sums, products, and contractions. Primitive invariants of more than two indices are traceless, since another invariant with less indices could otherwise be formed by contracting two indices of the first invariant. For simple groups, primitive invariants can be written in a form where they are either totally symmetric or totally antisymmetric~\cite{Fuchs:1997jv}. Coming back to the extended multi-Galileon Lagrangians, the first of them can be written as
\begin{equation}
\displaystyle{\mathcal{L}^{\text{ext}}_{_I} =  A_{\a\b\c\d\e}\delta^{\mu_1 \mu_2 \mu_3}_{\nu_1\nu_2\nu_3} \partial_{\mu_1}\pi^\a  
\partial_{\mu_2}\pi^\b\partial^{\nu_1}\pi^\c\partial^{\nu_2}\pi^\d \partial_{\mu_3}\partial^{\nu_3}\pi^\e,}
\end{equation}
with $A_{\a\b\c\d\e}$ built from primitive invariants. For example, in the fundamental representation of a SO($N$) symmetry group in a vector representation, the primitive invariants are only $\delta_{\a\b}$ and $\epsilon_{\au\ldots \aN}$. In the adjoint representation of a SU($N$) symmetry group described with one-tensors $\pi^\a$ for $\a \in \{1,\cdots,N^2-1\}$, the primitive invariants are $\delta_{\a\b}$, $f_{\a\b\c}$ the structure constants tensor, and $d_{\a\b\c}$ the completely symmetric invariant (for $N\geq 3$) \cite{Slansky:1981yr,Fuchs:1997jv,Ramond:2010zz}. These results will be discussed in detail in Secs.~\ref{PartFundRep} and~\ref{PartAdjRep}. In the following discussion, we will label the representations with only one internal index, to simplify the notation.

Note that when computing the equations of motion, there is a derivation with respect to a field. It means that the equations of motion will be in the conjugate representation of this field\footnote{We remind that the fundamental representation of the orthogonal groups and the adjoint representation of all simple groups are real and self-conjugate, while the fundamental representations of SU($N$) are complex-conjugate and in pair (for $N\geq3$) \cite{Slansky:1981yr}.}. So, the equations of motion will be linear combinations of terms with one free index in the primitive invariant part, that we will label $\al$, yielding
\begin{equation}
\displaystyle{\begin{array}{l}
EOM^{\text{ext}}_{_I}=  A^{\cdots\al\cdots} \delta^{\mu_1 \mu_2 \mu_3}_{\nu_1\nu_2\nu_3} \partial_{\mu_1}\pi \partial^{\nu_1}\pi\partial_{\mu_2}\partial^{\nu_2}\pi \partial_{\mu_3}\partial^{\nu_3}\pi\\
EOM^{\text{ext}}_{_{II}} =   A^{\cdots\al\cdots} \delta^{\mu_1 \mu_2 \mu_3 \mu_4}_{\nu_1\nu_2\nu_3\nu_4} \partial_{\mu_1}\pi \partial^{\nu_1}\pi\partial_{\mu_2}\partial^{\nu_2} \pi \partial_{\mu_3}\partial^{\nu_3}\pi\partial_{\mu_4}\partial^{\nu_4}\pi,\\
EOM^{\text{ext}}_{_{III}} =  A^{\cdots\al\cdots} \delta^{\mu_1 \mu_2 \mu_3 \mu_4}_{\nu_1\nu_2\nu_3\nu_4} \partial_{\mu_1}\pi  \partial_{\mu_2}\pi\partial^{\nu_1}\pi\partial^{\nu_2}\pi\partial_{\mu_3}\partial^{\nu_3}\pi\partial_{\mu_4}\partial^{\nu_4}\pi,
\end{array}}
\end{equation}
where we did not give explicitly the other contractions between group indices, and where in the whole paper the equations of motion read $EOM=0$. Note that the equations of motion for $\mathcal{L}^{\text{ext}}_{_I}$ and $\mathcal{L}^{\text{ext}}_{_{II}}$ do not reduce to multi-Galileon Lagrangians, since they are not in a singlet representation of the symmetry group.

To strengthen the examination of possible Lagrangians, we will also proceed to a systematic search of possible terms without using property~\textbf{b}. To study the possible Lagrangians that can be written, one has to consider the possible contractions between the independent primitive invariant prefactors and the Lagrangians given in Eq.~\eqref{EqExtGal}. Thanks to the symmetries of these Lagrangians and to property~\textbf{a}, only few possibilities will remain in most cases. This leads us to the following useful property:
\begin{itemize}
\item[\textbf{d)}] \textsl{If for a given primitive invariant prefactor, it is possible to write only one non-vanishing Lagrangian $\mathcal{L}_1$ which is not a total derivative by contracting this prefactor to a $\mathcal{L}^{\text{ext}}_i$, and if it is possible to write a nonvanishing current by taking off one of the derivatives of a second-order derivative contribution in this Lagrangian, then this Lagrangian is a total derivative.}
\end{itemize}
To prove this result, it is sufficient to consider that the total derivative obtained when taking the divergence of the previously formed current only reduces to contractions of the given primitive invariant prefactor with $\mathcal{L}^{\text{ext}}_i$. Thus, the divergence of this conserved current will give $\mathcal{L}_1$ and Lagrangians which are either vanishing or total derivatives. It shows that $\mathcal{L}_1$ is also a total derivative. Using the same method, one can show the following property:{Let us call $J^\mu_{1}$ one of these currents associated to $\mathcal{L}_{1}$. Its divergence gives $\mathcal{L}_{1}$ and additional Lagrangians. Either all these Lagrangians vanish or are total derivatives, or they contain $\mathcal{L}_{2}$. In the first case, $\mathcal{L}_{1}$ is a total derivative, in the second case, $\mathcal{L}_{1}$ implies the same dynamics as $\mathcal{L}_{2}$.}
\begin{itemize}
\item[\textbf{e)}] \textsl{If for a given primitive invariant prefactor, it is possible to write only two non-vanishing Lagrangians $\mathcal{L}_j$ which are not total derivatives by contracting this prefactor with a $\mathcal{L}^{\text{ext}}_i$, and if it is possible to write a nonvanishing current by taking off one of the derivatives of a second-order derivative term of one of these Lagrangians, then they imply at most one nontrivial dynamics.}
\end{itemize}
Both these properties will be useful in the following.

\section{Extended Multi-Galileons in SO($N$) and SU($N$) fundamental representations}
\label{PartFundRep}
\subsection{SO($N$) fundamental representation}
\label{PartIntroSON}

In this section, we consider real multi-Galileon fields which transform in the fundamental representation of a SO($N$) global symmetry group. They are labeled as vectors with group indices going from $1$ to $N$. The action of the group on them corresponds to the defining rotation matrices of SO($N$):
\begin{equation}
\pi^\a \rightarrow O^{\a}{}_{\b} \pi^{\b}.
\end{equation}
The representation being self-conjugate and real, the upper and lower indices are equivalent, and can be exchanged with the flat metric $\delta_{\a\b}$. The only primitive invariants of the orthogonal group in this representation are the Kronecker delta $\delta_{\a\b}$ and the Levi-Civita tensor $\epsilon_{\au\cdots\aN}$. 

Following property~\textbf{b} and keeping in mind the Lagrangians of Eq.~\eqref{EqExtGal}, it is necessary to build from the primitive invariants totally symmetric terms of rank 3 for $\mathcal{L}^{\text{ext}}_{_I}$ and $\mathcal{L}^{\text{ext}}_{_{III}}$, and of rank 4 for $\mathcal{L}^{\text{ext}}_{_{II}}$. It is only possible for rank 4, yielding $\delta_{(\a\b}\delta_{\c\d)}$, where the parentheses denote a symmetrization with the following normalization
\begin{equation}
\label{EqNormDeltaSON}
\delta_{(\a\b}\delta_{\c\d)} \equiv \delta_{\a\b}\delta_{\c\d} + \delta_{\a\c}\delta_{\b\d}+ \delta_{\a\d}\delta_{\b\c}.
\end{equation}
Thus, the only viable dynamics in this representation is described by the Lagrangian
\begin{equation}
\mathcal{L}^{\text{SO}(N)} = \delta^{\mu_1 \mu_2 \mu_3 \mu_4}_{\nu_1\nu_2\nu_3\nu_4} \partial_{\mu_1}\pi^\a  \partial_{\mu_2}\pi^\b\partial^{\nu_1}\pi^\c\partial^{\nu_2}\pi^\d \partial_{\mu_3}\partial^{\nu_3}\pi^\e\partial_{\mu_4}\partial^{\nu_4}\pi^\f \delta_{(\a\c}\delta_{\e\f)}\delta_{\b\d}.
\label{EqLagUniSON}
\end{equation}
Note that as expected, the invariant prefactor is also symmetric under the exchange $\b\leftrightarrow\d$. Its equation of motion yields
\begin{multline}
\displaystyle{EOM^{\text{SO}(N)} = 8 \delta^{\mu_1 \mu_2 \mu_3 \mu_4}_{\nu_1\nu_2\nu_3\nu_4}
\left[ \partial_{\mu_1}\pi^\b \partial^{\nu_1}\pi^\al\partial_{\mu_2}\partial^{\nu_2}\pi_\b \partial_{\mu_3}\partial^{\nu_3}\pi^\c\partial_{\mu_4}\partial^{\nu_4}\pi_\c 
\right. }\\\displaystyle{ \left. 
- 2  \partial_{\mu_1}\pi^\b \partial^{\nu_1}\pi_\b \partial_{\mu_2}\partial^{\nu_2}\pi^\al \partial_{\mu_3}\partial^{\nu_3}\pi^\c\partial_{\mu_4}\partial^{\nu_4}\pi_\c
+ \partial_{\mu_1}\pi^\b \partial^{\nu_1}\pi^\c\partial_{\mu_2}\partial^{\nu_2}\pi^\al \partial_{\mu_3}\partial^{\nu_3}\pi_\b\partial_{\mu_4}\partial^{\nu_4}\pi_\c\right],}
\label{EqEOMSON}
\end{multline}
in its simpler form, where $\al$ is the free index denoting the fact that this equation is in the fundamental representation of the SO($N$) symmetry group. Finally, one can verify that 
\begin{equation}
\displaystyle{EOM^{\text{SO}(N)}  = 8 ~ \partial_{\mu_2} \left[J_1^{\mu_2} - J_2^{\mu_2} \right],}
\end{equation}
with 
\begin{equation}
\displaystyle{
\begin{array}{l}
J_1^{\mu_2} = \delta^{\mu_1 \mu_2 \mu_3 \mu_4}_{\nu_1\nu_2\nu_3\nu_4} \partial_{\mu_1}\pi^\b \partial^{\nu_1}\pi^\al\partial^{\nu_2}\pi_\b \partial_{\mu_3}\partial^{\nu_3}\pi^\c\partial_{\mu_4}\partial^{\nu_4}\pi_\c,\\
J_2^{\mu_2} = \delta^{\mu_1 \mu_2 \mu_3 \mu_4}_{\nu_1\nu_2\nu_3\nu_4} \partial_{\mu_1}\pi^\b \partial^{\nu_1}\pi_\b \partial^{\nu_2}\pi^\c \partial_{\mu_3}\partial^{\nu_3}\pi^\al\partial_{\mu_4}\partial^{\nu_4}\pi_\c,
\end{array}}
\end{equation}
as expected from property~\textbf{c}.

\subsection{Exhaustive examination and alternative formulations}

In this section, we make an exhaustive investigation of all the possible Lagrangian terms for multi-Galileons in the fundamental representation of a SO($N$) symmetry group, i.e. without using property~\textbf{b}. For that purpose, we consider the prefactors built from the primitive invariants of this representation that give nonvanishing contractions with the $\mathcal{L}^{\text{ext}}_j$ Lagrangians of Eq.~\eqref{EqExtGal}. It allows us to verify that the Lagrangian introduced in Sec.~\ref{PartIntroSON} describes the only possible dynamic for extended multi-Galileons in this representation. It also allows us to describe in more detail the alternative formulations of the model and their properties. 

Let us first consider the contractions with $\delta_{\a\b}$ only. These contractions are only possible with $\mathcal{L}^{\text{ext}}_{_{II}}$, since it is the only Lagrangian with an even number of fields. Then, taking into account the symmetries of this Lagrangian, the only two independent non-vanishing contractions are 
\begin{equation}
\displaystyle{
\begin{array}{l}
\mathcal{L}^{\text{SO}(N)}_1 = \delta^{\mu_1 \mu_2 \mu_3 \mu_4}_{\nu_1\nu_2\nu_3\nu_4} \partial_{\mu_1}\pi^\a  \partial_{\mu_2}\pi^\b\partial^{\nu_1}\pi_\a\partial^{\nu_2}\pi_\b \partial_{\mu_3}\partial^{\nu_3}\pi^\c\partial_{\mu_4}\partial^{\nu_4}\pi_\c,\\
\mathcal{L}^{\text{SO}(N)}_2 = \delta^{\mu_1 \mu_2 \mu_3 \mu_4}_{\nu_1\nu_2\nu_3\nu_4} \partial_{\mu_1}\pi^\a  \partial_{\mu_2}\pi^\b\partial^{\nu_1}\pi^\c\partial^{\nu_2}\pi_\b \partial_{\mu_3}\partial^{\nu_3}\pi_\a\partial_{\mu_4}\partial^{\nu_4}\pi_\c.\\
\end{array}}
\label{EqLagEqSON}
\end{equation}
Following property~\textbf{e}, it is straightforward to show that they are related by a total derivative. Indeed, taking 
\begin{equation}
\displaystyle{J^{\mu_4}_{1\leftrightarrow2} = \delta^{\mu_1 \mu_2 \mu_3 \mu_4}_{\nu_1\nu_2\nu_3\nu_4} \partial_{\mu_1}\pi^\a  \partial_{\mu_2}\pi^\b\partial^{\nu_1}\pi_\a\partial^{\nu_2}\pi_\b \partial_{\mu_3}\partial^{\nu_3}\pi^\c\partial^{\nu_4}\pi_\c,}
\end{equation}
we get
\begin{equation}
\label{EqSONCurrent}
\displaystyle{\partial_{\mu_4} J^{\mu_4}_{1\leftrightarrow2} = \mathcal{L}^{\text{SO}(N)}_1 - 2 \mathcal{L}^{\text{SO}(N)}_2.}
\end{equation}
In addition, these two Lagrangians can be related to $\mathcal{L}^{\text{SO}(N)}$, yielding
\begin{equation}
\mathcal{L}^{\text{SO}(N)} = \mathcal{L}^{\text{SO}(N)}_1 + 2 \mathcal{L}^{\text{SO}(N)}_2.
\label{EqSONSum}
\end{equation}
Using Eqs.~\eqref{EqSONCurrent} and~\eqref{EqSONSum}, we see that all three Lagrangians imply the same dynamics. These alternative Lagrangians illustrate the discussion following property~\textbf{b}, since they have a nontrivial dynamic and do not explicitly have the set of four indices which are totally symmetric under the exchange described in property~\textbf{b}. However, this symmetry can be recovered as expected by adding the total derivative of a current with a set of three indices that are totally symmetric under exchange, i.e. the lower indices $\a$, $\b$ and $\c$ of the current $J^{\mu_4}_{1\leftrightarrow2}$. One can note that the most symmetric version of the Lagrangians is not necessarily the simplest one. For example, the equation of motion given in Eq.~\eqref{EqEOMSON} is easier to obtain from the alternative formulations given in Eq.~\eqref{EqLagEqSON}.

The Lagrangians that can be constructed with primitive invariant prefactors containing antisymmetric Levi-Civita tensors are examined in Appendix.~\ref{AppEpsSON}. None of them give a nontrivial dynamics, as expected. This concludes the complete investigation of possible Lagrangian terms for extended multi-Galileons in the fundamental representation of a SO($N$) global symmetry group, showing that only one dynamics is possible, described by the Lagrangian of Eq.~\eqref{EqLagUniSON}. 

\subsection{SU($N$) fundamental representation}

We discuss here the case of complex multi-Galileon fields which transform in the fundamental representation of a SU($N$) global symmetry group. In this case, we have to consider also its complex-conjugate representation. For SU(2), both representations are equivalent, but the fundamental representation is pseudoreal. Then, there exists no basis in which the action of the group elements on the fundamental representation and its complex-conjugate representation are equal, and it is thus better to distinguish them. For SU($N$) with $N\geq3$, which we focus on from now on, the complex-conjugate representations are inequivalent, and have to be distinguished.

We use vector notations with upper indices for the fundamental representation, labeling the multi-Galileon fields by $\pi^\a$ with $\a$ from $1$ to $N$. The complex-conjugates of these fields are labeled with lower indices, i.e. denoting  $\pi_\a = \left(\pi^\a\right)^*$. With these notations, they transform under the action of the defining matrices of SU($N$) as 
\begin{equation}
\pi^\a \rightarrow U^{\a}{}_{\b} \pi^\b, ~~~ ~~~ ~~~ \pi_\a \rightarrow U^{\b}{}_{\a} \pi_\b.
\end{equation}
It is important to keep in mind that lower and upper indices label two different representations, and cannot be exchanged by the application of a group metric, as in the SO($N$) case. There are three primitive invariants in this representation, the Kronecker delta $\delta^{\a}{}_{\b}$ and two Levi-Civita tensors $\epsilon_{\au\cdots\aN}$ and $\epsilon^{\au\cdots\aN}$.

To obtain the possible nontrivial Lagrangians satisfying property~\textbf{b} in this case, one has to consider the possible totally symmetric terms built from primitive invariants of ranks 3 and 4. As in the SO($N$) case, it is only possible to build such a term for rank 4, by symmetrizing two Kronecker deltas. This symmetrization has to be done independently of the representations of the fields. The symmetric tensor thus yields
\begin{equation}
\displaystyle{
\delta^{(\a}{}_{\b} \delta^{\c}{}_{\d)} \equiv \delta^{\a}{}_{\b} \delta^{\c}{}_{\d} + \delta^{\a}{}_{\b} \delta^{\d}{}_{\c}  + \delta^{\a}{}_{\c} \delta^{\b}{}_{\d} + \delta^{\a}{}_{\c} \delta^{\d}{}_{\b} + \delta^{\a}{}_{\d} \delta^{\b}{}_{\c} + \delta^{\a}{}_{\d} \delta^{\c}{}_{\b}.
}
\end{equation}
In this equation, one has to pay attention to the fact that the lower indices have to be contracted with $\pi^\a$ fields, while the upper indices have to be contracted with complex-conjugate $\pi_\a$ fields. Here, the alphabetical order merely is an indication of the position of the field in the Lagrangian. Then, the only possible Lagrangian is 
\begin{equation}
\displaystyle{
\mathcal{L}^{\text{SU}(N)} = \delta^{\mu_1 \mu_2 \mu_3 \mu_4}_{\nu_1\nu_2\nu_3\nu_4} \partial_{\mu_1}\pi_\a  \partial_{\mu_2}\pi_\b\partial^{\nu_1}\pi^\c\partial^{\nu_2}\pi^\d \partial_{\mu_3}\partial^{\nu_3}\pi_\e\partial_{\mu_4}\partial^{\nu_4}\pi^\f \delta^{(\a}{}_{\c} \delta^{\e}{}_{\f)} \delta^{\b}{}_{\d} +  \text{h.c.},
}
\label{EqLagSUNFund}
\end{equation}
whose equations of motion are similar but more involved than the equations of motion of the real case.

Looking for alternative formulations, one has to pay attention that it is a priori possible to build more Lagrangians in the SU($N$) case than in the SO($N$) case. First, the necessity to distinguish fields with upper and lower indices allows us to write several Lagrangians which reduce to the ones of the SO($N$) case in the limit where the fields become real (i.e. with a real representation in which complex-conjugate fields are equivalent). For example, the two Lagrangians\footnote{
Note that here, $\mathcal{L}^{\text{SU}(N)}_1$ to $\mathcal{L}^{\text{SU}(N)}_3$ are written with expressions which are already real, thanks to their symmetry properties. We however prefer to write them explicitly with the h.c. contribution, to remind the reader that they are built from complex-conjugate fields. In addition, it simplifies the calculations when dealing with other Lagrangians which do not have this property, as it is the case e.g. for
\begin{equation}
\displaystyle{
\mathcal{L}^{\text{SU}(N)}_4 =  \delta^{\mu_1 \mu_2 \mu_3 \mu_4}_{\nu_1\nu_2\nu_3\nu_4} \partial_{\mu_1}\pi^\a  \partial_{\mu_2}\pi^\b\partial^{\nu_1}\pi_\b\partial^{\nu_2}\pi^\c \partial_{\mu_3}\partial^{\nu_3}\pi_\a\partial_{\mu_4}\partial^{\nu_4}\pi_\c + \text{h.c.}.
}
\end{equation}
}
\begin{equation}
\displaystyle{
\begin{array}{l}
\mathcal{L}^{\text{SU}(N)}_1 =  \delta^{\mu_1 \mu_2 \mu_3 \mu_4}_{\nu_1\nu_2\nu_3\nu_4} \partial_{\mu_1}\pi^\a  \partial_{\mu_2}\pi^\b\partial^{\nu_1}\pi_\a\partial^{\nu_2}\pi_\b \partial_{\mu_3}\partial^{\nu_3}\pi^\c\partial_{\mu_4}\partial^{\nu_4}\pi_\c + \text{h.c.},\\
\mathcal{L}^{\text{SU}(N)}_2 =  \delta^{\mu_1 \mu_2 \mu_3 \mu_4}_{\nu_1\nu_2\nu_3\nu_4} \partial_{\mu_1}\pi^\a \partial_{\mu_2}\pi_\b\partial^{\nu_1}\pi_\a\partial^{\nu_2}\pi^\b \partial_{\mu_3}\partial^{\nu_3}\pi^\c\partial_{\mu_4}\partial^{\nu_4}\pi_\c + \text{h.c.},
\end{array}
}
\end{equation}
reduce to $\mathcal{L}^{\text{SO}(N)}_1$ of Eq.~\eqref{EqLagEqSON} in the real case. Similarly, three different Lagrangians can be written in the SU($N$) case which reduce to $\mathcal{L}^{\text{SO}(N)}_2$ of Eq.~\eqref{EqLagEqSON}. Then, dealing with two representations which are not equivalent, it is also possible to write down Lagrangians which would vanish due to symmetry considerations in the SO($N$) limit. One can for example consider
\begin{equation}
\displaystyle{
\mathcal{L}^{\text{SU}(N)}_3 =  \delta^{\mu_1 \mu_2 \mu_3 \mu_4}_{\nu_1\nu_2\nu_3\nu_4} \partial_{\mu_1}\pi^\a  \partial_{\mu_2}\pi_\a\partial^{\nu_1}\pi_\b\partial^{\nu_2}\pi^\c \partial_{\mu_3}\partial^{\nu_3}\pi^\b\partial_{\mu_4}\partial^{\nu_4}\pi_\c + \text{h.c.},
}
\end{equation}
which identically vanishes in the real limit. This indicates that the alternative formulations of $\mathcal{L}^{\text{SU}(N)}$ are a priori more numerous than in the real case.

\subsection{Summary}

For the extended multi-Galileon models in the fundamental representation of a global SO($N$) symmetry group, the only nontrivial dynamics is given by 
\begin{equation}
\label{EqLagFinSON}
\mathcal{L}^{\text{SO}(N)} = \delta^{\mu_1 \mu_2 \mu_3 \mu_4}_{\nu_1\nu_2\nu_3\nu_4} \partial_{\mu_1}\pi^\a  \partial_{\mu_2}\pi^\b\partial^{\nu_1}\pi^\c\partial^{\nu_2}\pi^\d \partial_{\mu_3}\partial^{\nu_3}\pi^\e\partial_{\mu_4}\partial^{\nu_4}\pi^\f \delta_{(\a\c}\delta_{\e\f)}\delta_{\b\d}.
\end{equation}
Equivalent Lagrangians are given in Eq.~\eqref{EqLagEqSON}, and its equation of motion can be found in Eq.~\eqref{EqEOMSON}. There is also one possible nontrivial dynamics in the case of the fundamental representation of a global SU($N$) symmetry group, where the only possible dynamics is described by the Lagrangian given in Eq.~\eqref{EqLagSUNFund}. However, more equivalent Lagrangians can a priori be written in this case.

This whole section showed that as discussed previously, the properties of the extended multi-Galileon models are very constraining and allow us to build only a few independent nontrivial dynamics, even if it is possible to write a lot of nonvanishing Lagrangians. Properly used, these properties drastically simplify the research of possible nontrivial Lagrangian terms. 

\section{Extended multi-Galileons in the SU($N$) adjoint representation}
\label{PartAdjRep}
\subsection{SU($N$) adjoint representation}
\label{PartIntroSUN}

In this section, we consider multi-Galileon fields in the adjoint representation of a global SU($N$) symmetry group. We will denote these fields with one single index going from 1 to $N^2-1$ as the dimension of the group (i.e. using the adjoint module of the group). The action of the group elements labeled with $\a$ on a multi-Galileon $\pi_\b$ yields
\begin{equation}
\pi_\b ~~~ \rightarrow ~~~ (T_\a)_\b{}^\c \pi_\c = f_{\a\b}{}^\c \pi_\b,
\end{equation}
where $f_{\a\b}{}^\c$ the structure constants of SU($N$). We chose here to focus on the adjoint representation of a SU($N$) rather than a SO($N$) symmetry group, due to the fact that the low-rank SO($N$) groups can be related to other symmetry groups thanks to local isomorphisms. This is for example the case for SO(3) and SU(2), SO(4) and SU(2)$\times$SU(2), SO(5) and Sp(4), and SO(6) and SU(4).

The metric on this representation is $g_{\a\b}=f_{\a\al}{}^\be f_{\b\be}{}^\al$, and can be used to raise and lower the indices. It allows us to work with the completely antisymmetric forms of the structure constant, $f_{\a\b\c}$, as well as $f^{\a\b\c}$. The primitive invariants in this representation are $g_{\a\b}$, $f_{\a\b\c}$, and $d_{\a\b\c}$ which is the symmetric primitive invariant for $N\geq3$ \cite{Fuchs:1997jv,deAzcarraga:1997ya,Ramond:2010zz}. In the SU(3) case, the link between this basis and the one of the defining matrices of SU($N$) was discussed e.g. in Ref.~\cite{Dittner:1972hm}. We choose a basis where the group metric is proportional to the Kronecker delta $\delta_{\a\b}$. In addition, we can impose the following normalizations \cite{Fuchs:1997jv}
\begin{equation}
f^{\a\b\c}f^{\a\b\d}= N \delta^{\c\d}, ~~~ ~~~ d^{\a\b\c}d^{\a\b\d} = \left(N - \frac4N \right) \delta^{\c\d}.
\end{equation}
The values of $f_{\a\b\c}$ and $d_{\a\b\c}$ in the SU(3) case can be found e.g. in Ref.~\cite{deAzcarraga:1997ya}. Concerning the contractions of these primitive invariants, some properties are very useful. First, the contractions of primitive invariants with the Kronecker delta tensor can be omitted since this tensor only raises or lowers the group indices. Then, the contractions between $f$ and $d$ primitive invariants of ranks zero to three, i.e. with zero to three indices not contracted together, are also known~\cite{Metha:1983mng,Fuchs:1997jv}; they vanish for ranks zero and one, are proportional to the Kronecker delta for rank two, and proportional to $f$ or $d$ for rank three\footnote{The result is proportional to $f$ if there is an odd number of $f$ and an even number of $d$, and to $d$ if there is an even number of $f$ and an odd number of $d$. The case where the numbers of $f$ and $d$ have the same parity is not possible since it would not be possible to have three indices not contracted.}. 

Following property~\textbf{b}, we will discuss the possible independent dynamics by considering the totally symmetric terms of ranks 3 and 4 that can be built from primitive invariants. It will allow us to build only a few independent Lagrangians, that we will examine in Sec.~\ref{PartSUNLag}. In Appendix~\ref{AppPartSUN}, we also make an exhaustive examination of all the possible Lagrangians at the order of $\mathcal{L}^{\text{ext}}_{_I}$. Finding similar Lagrangians by both methods corroborates our approach, and shows that we effectively describe all the possible dynamics for multi-Galileon fields in the adjoint representation of a SU($N$) symmetry group. The cases of SU(2) and SU(3) symmetry groups  are then discussed in Sec.~\ref{PartSU2&3}. 

\subsection{Multi-Galileon Lagrangians}
\label{PartSUNLag}

To obtain the Lagrangians implying the different nontrivial dynamics of this model, we need to discuss the possible totally symmetric invariants of ranks 3 and 4. For rank 3, only one such invariant is possible, $d_{\a\b\c}$, which is nonzero only for $N\geq3$. For rank 4, two such invariants are possible. The first one can be built from the Kronecker delta, and gives $\delta_{(\a\b}\delta_{\c\d)}$, where we take the normalization of Eq.~\eqref{EqNormDeltaSON}. The second one is the Casimir invariant at order $4$, for $N\geq 4$ (see e.g. Refs.~\cite{Fuchs:1997jv,Ramond:2010zz}), and can be written as
\begin{equation}
\label{PrimInvFour}
\displaystyle{d_{\be(\a\b} d^{\be}{}_{\c\d)} \equiv d_{\be\a\b} d^{\be}{}_{\c\d} + d_{\be\a\c} d^{\be}{}_{\b\d} +d_{\be\a\d} d^{\be}{}_{\b\c}.}
\end{equation}
With the knowledge of these totally symmetric invariants, one can thus investigate the possible Lagrangians with a nontrivial dynamics, using property~\textbf{b} as well as the form of the Lagrangians given in Eq.~\eqref{EqExtGal}. 

At the order of $\mathcal{L}^{\text{ext}}_{_I}$, only one Lagrangian is possible, yielding
\begin{equation}
\label{LagFinSUN1}
\displaystyle{\mathcal{L}^{\text{Adj},1}_{_I} =  \delta^{\mu_1 \mu_2 \mu_3}_{\nu_1\nu_2\nu_3} \partial_{\mu_1}\pi^\a  
\partial_{\mu_2}\pi^\b\partial^{\nu_1}\pi^\c\partial^{\nu_2}\pi^\d \partial_{\mu_3}\partial^{\nu_3}\pi^\e d_{\a\c\e}\delta_{\b\d}.}
\end{equation}
Note that a systematic examination of all the possible terms at this order is presented in Appendix~\ref{AppPartSUN}, showing that the dynamics implied by this Lagrangian is the only nontrivial one. The equations of motion of this Lagrangian are given in Eq.~\eqref{AppEqEOMSUN}.
At the order of $\mathcal{L}^{\text{ext}}_{_{II}}$, the two possible fourth-rank totally symmetric invariants can be used. It yields the two following Lagrangians
\begin{equation}
\label{LagFinSUN2}
\displaystyle{
\begin{array}{l}
\mathcal{L}^{\text{Adj},1}_{_{II}} =  \delta^{\mu_1 \mu_2 \mu_3 \mu_4}_{\nu_1\nu_2\nu_3\nu_4} \partial_{\mu_1}\pi^\a  \partial_{\mu_2}\pi^\b\partial^{\nu_1}\pi^\c\partial^{\nu_2}\pi^\d \partial_{\mu_3}\partial^{\nu_3}\pi^\e\partial_{\mu_4}\partial^{\nu_4}\pi^\f \delta_{(\a\c}\delta_{\e\f)}\delta_{\b\d},\\
\mathcal{L}^{\text{Adj},2}_{_{II}} =  \delta^{\mu_1 \mu_2 \mu_3 \mu_4}_{\nu_1\nu_2\nu_3\nu_4} \partial_{\mu_1}\pi^\a  \partial_{\mu_2}\pi^\b\partial^{\nu_1}\pi^\c\partial^{\nu_2}\pi^\d \partial_{\mu_3}\partial^{\nu_3}\pi^\e\partial_{\mu_4}\partial^{\nu_4}\pi^\f d_{\be(\a\c} d^{\be}{}_{\e\f)}\delta_{\b\d}.
\end{array}
}
\end{equation}
At the order of $\mathcal{L}^{\text{ext}}_{_{III}}$, one needs a rank-three symmetric tensor, which can only be $d_{\a\b\c}$. Then, four indices still have to be contracted, yielding the following possibilities:
\begin{equation}
\label{LagFinSUN3}
\displaystyle{
\begin{array}{l}

\mathcal{L}^{\text{Adj},1}_{_{III}} =  \delta^{\mu_1 \mu_2 \mu_3 \mu_4}_{\nu_1\nu_2\nu_3\nu_4} \partial_{\mu_1}\pi^\a  \partial_{\mu_2}\pi^\b\partial_{\mu_3}\pi^\c\partial^{\nu_1}\pi^\d\partial^{\nu_2}\pi^\e\partial^{\nu_3}\pi^\f \partial_{\mu_4}\partial^{\nu_4}\pi^\g d_{\a\d\g} \delta_{\b\e}\delta_{\c\f}.\\
\mathcal{L}^{\text{Adj},2}_{_{III}} =  \delta^{\mu_1 \mu_2 \mu_3 \mu_4}_{\nu_1\nu_2\nu_3\nu_4} \partial_{\mu_1}\pi^\a  \partial_{\mu_2}\pi^\b\partial_{\mu_3}\pi^\c\partial^{\nu_1}\pi^\d\partial^{\nu_2}\pi^\e\partial^{\nu_3}\pi^\f \partial_{\mu_4}\partial^{\nu_4}\pi^\g d_{\a\d\g} d^{\be}{}_{\b\e} d_{\be\d\f},
\end{array}
}
\end{equation}
There are possible contractions with terms in $ d_{\a\d\g}f^{\be}{}_{\b\c} f_{\be\e\f}$ and $d_{\a\d\g}f^{\be}{}_{\b\e} f_{\be\c\f}$, which are related by the Jacobi identity. But these terms would in fact identically vanish, since they imply an antisymmetrical contraction between $\a$ and $\d$. For example, exchanging the positions of $\b$ and $\e$ thanks to a $f^{\be}{}_{\b\e}$ terms shows the antisymmetry between $\a$ and $\e$. Then, and starting from the initial configuration, exchanging the positions of $\a$ and $\e$ thanks to the previous antisymmetry puts the Lagrangian in a form which involves a $\partial^{\nu_1} \pi^{\d} \partial^{\nu_2} \pi^{\a}$ part which identically vanishes due to the contractions with $\delta^{\mu_1 \mu_2 \mu_3 \mu_4}_{\nu_1\nu_2\nu_3\nu_4}$ and $d_{\a\d\g}$. 

Finally, we exactly recover the symmetries of property~\textbf{b}. Indeed, the Lagrangians built from $\mathcal{L}^{\text{ext}}_{_I}$ and $\mathcal{L}^{\text{ext}}_{_{II}}$ are symmetric under the exchange $\b\leftrightarrow\d$, and the Lagrangians built from $\mathcal{L}^{\text{ext}}_{_{III}}$ are symmetric by pairs for the set of indices $(\b,\c,\e,\f)$. These Lagrangians describe the only possible dynamics of extended multi-Galileon models with fields in the adjoint representation of a SU($N$) symmetry group. The model is simplified in the case of $N=2$ or $N=3$, as discussed in Sec.~\ref{PartSU2&3}. Note also that the properties we introduced drastically simplify the investigation of such a model. For example, investigating $\mathcal{L}^{\text{ext}}_{_{III}}$ would otherwise imply considering the 1557 singlet configurations built from seven adjoint fields in the case of SU(4) ~\cite{Slansky:1981yr,Feger:2012bs}.

\subsection{SU(2) and SU(3) cases}
\label{PartSU2&3}

In the previous section, we investigated the case of a general SU($N$) symmetry group. However, this study can be simplified in the case of a SU(3) or SU(2) symmetry group. In the case of SU(3), there is only one fourth-rank Casimir symmetric invariant. Indeed, one has the following relations between the primitive invariants~\cite{Fuchs:1997jv}:
\begin{equation}
 d_{\a\b\be}d_{\c\d\be} = \frac12\left(\delta_{\a\c}\delta_{\b\d} + \delta_{\b\c}\delta_{\a\d} - \delta_{\a\b}\delta_{\c\d} \right) +  f_{\a\c\be}f_{\b\d\be} +  f_{\a\d\be}f_{\b\c\be} ,
\end{equation}
allowing us to write as expected $d_{\be(\a\b} d^{\be}{}_{\c\d)}$ as a function of the Kronecker deltas and terms implying the structure constant tensors which are total derivative contributions thanks to property \textbf{a}. The same relation can be used to write the Lagrangian $\mathcal{L}^{\text{Adj},2}_{_{III}}$ as a linear combination of $\mathcal{L}^{\text{Adj},1}_{_{III}}$ plus some terms which give identically vanishing Lagrangians as explained in the previous section. Thus, the Lagrangians which describe the possible independent dynamics for extended multi-Galileon models in the adjoint representation of a SU(3) symmetry group are $\mathcal{L}^{\text{Adj},1}_{_I}$, $\mathcal{L}^{\text{Adj},1}_{_{II}}$ and $\mathcal{L}^{\text{Adj},1}_{_{III}}$.

The case of SU(2) is even simpler. In this symmetry group, there is no $d_{\a\b\c}$ primitive invariant. Therefore, most of the Lagrangians of the general SU($N$) case vanish. The only possible dynamics is thus described by the $\mathcal{L}^{\text{Adj},1}_{_{II}}$ Lagrangian. One can note that as SU(2) is locally isomorphic to SO(3), one should recover the same results for the three-dimensional representations of both groups. It is indeed the case, since the result for the adjoint representation of SU(2) is exactly the one we found for the fundamental representation of SO(3) in Sec.~\ref{PartFundRep}.

\section{Conclusion and discussions}
\label{PartConclusion}

In this paper, we investigated possible terms included in the generalized multi-Galileon theories, which we call extended multi-Galileon Lagrangians. Some of these terms were already introduced in the literature: the possibility for more than two fields with first-order derivative only to be contracted with the same $\delta^{\mu_1\cdots\mu_m}_{\nu_1\cdots\nu_m}$ was discussed in Ref.~\cite{Deffayet:2010zh} for a theory containing several $p$-forms, and the Lagrangian $\mathcal{L}^{\text{ext}}_{_I}$ for biscalar theories was given in Ref.~\cite{Ohashi:2015fma}. We performed here a systematic examination of those Lagrangians. We first discussed the general properties of these Lagrangians, showing that they are strongly constrained by symmetry relations, which can be explicit or hidden up to a total derivative. Then, using these symmetry properties, we examined the possible dynamics for multi-Galileons in the fundamental representation of a SO($N$) or SU($N$) symmetry group, and for the adjoint representation of a SU($N$) symmetry group. In the case of the fundamental representation of SO($N$) and of certain terms of the adjoint of SU($N$), we also performed in parallel a complete investigation of the possible nontrivial dynamics. The results of these investigations are in complete agreement with the Lagrangians built from symmetry considerations, and also allowed us to discuss internal properties of the model.

A next step would be to investigate what would be the most general theory for multi-Galileon fields. Following the works of Refs.~\cite{Padilla:2010de,Padilla:2010tj,Sivanesan:2013tba,Padilla:2012dx}, we would expect Lagrangians of the form 
\begin{multline}
\displaystyle{
\mathcal{L}= A^{\color{red}a_1 b_1  \color{black}\hspace{-0.05cm}\cdots  \color{red}a_n b_n c_1 \color{black}\hspace{-0.05cm}\cdots \color{red}c_{(m-n)}\color{black}}\left(X_{\a\b},\pi_\c \right) 
\delta^{\mu_1 \cdots \mu_m}_{\nu_1\cdots\nu_m} 
\partial_{\mu_1} \pi_{\color{red}a_1\color{black}} \partial^{\nu_1} \pi_{\color{red}b_1\color{black}} \hspace{-0.15cm}\cdots \partial_{\mu_n} \pi_{\color{red}a_n\color{black}} \partial^{\nu_n} \pi_{\color{red}b_n\color{black}}
}\\\displaystyle{
\partial_{\mu_{(n+1)}}\partial^{\nu_{(n+1)}} \pi_{\color{red}c_1\color{black}}
\hspace{-0.15cm}\cdots
\partial_{\mu_{m}}\partial^{\nu_{m}} \pi_{\color{red}c_{(m-n)}\color{black}}
 }
\end{multline}
where $X_{\a\b}= (1/2) \partial_\rho \pi_\a \partial^\rho \pi_\b$, $m$ goes from 1 to 4, and $n$ from 0 to $m-1$, and where we also expect $\partial A^{\color{red}a_1 b_1  \color{black}\hspace{-0.05cm}\cdots  \color{red}a_n b_b c_1 \color{black}\hspace{-0.05cm}\cdots \color{red}c_{(m-n)}\color{black}} /\partial X_{\a\b}$ to be symmetric in its indices $(\color{red}c_1\color{black},\cdots,\color{red}c_{(m-n)}\color{black},\a,\b)$ in order to have second-order equations of motion. An investigation of these Lagrangians for multi-Galileons in group representations of a global symmetry group would also allow us to have a better understanding of their properties. It would then be interesting to investigate this kind of models while allowing the theory to be degenerate, with Lagrangians potentially involving third-order-derivatives equations of motion, i.e. in a beyond-Horndeski context~\cite{Gleyzes:2014dya,Langlois:2015cwa,Langlois:2015skt,Motohashi:2016ftl,Crisostomi:2016czh}.

The link with vector multi-Galileon models can also be very fruitful. Vector Galileon models have been examined in Refs.~\cite{Heisenberg:2014rta,Tasinato:2014eka,Allys:2015sht,Jimenez:2016isa,Allys:2016jaq}, and their cosmological applications have been explored e.g. in Refs.~\cite{Tasinato:2014eka,Tasinato:2014mia,Hull:2014bga,DeFelice:2016cri,DeFelice:2016yws,DeFelice:2016uil,Heisenberg:2016wtr}. Those models are built from the same requirements as those of scalar Galileons, in addition to the requirement that the vector field propagate at most three degrees of freedom. Its longitudinal component described by considering the pure scalar part of the vector only, i.e. considering the $\partial_\mu \pi$ contribution to $A_\mu$ in its scalar-vector decomposition, should also have second-order equations of motion.

Recently, models of vector multi-Galileons have been developed \cite{Allys:2016kbq,Jimenez:2016upj}. The case of vector Galileons lying in the adjoint representation of a SU(2) global symmetry is particularly interesting since it can source or contribute to the inflation while keeping isotropy \cite{Golovnev:2008cf,ArmendarizPicon:2004pm}. In those models, vector multi-Galileons are denoted $A_\mu^\a$, and transform for the group indices similarly to the scalar multi-Galileons, see Sec.~\ref{PartIntroSUN}. Starting from a vector Lagrangian, and considering its pure longitudinal contribution, i.e. doing the replacement $A_\mu^\a \rightarrow \partial_\mu \pi^\a$, one recovers a scalar multi-Galileon model. In fact, this link is possible for the vector Lagrangians whose derivative parts can be written as a function of its symmetric form $S_{\mu\nu}^\a = \partial_\mu A_\nu^\a + \partial_\nu A_\mu^\a$ only, the antisymmetric one vanishing in the pure scalar sector.

Thus, it is also possible to obtain vector Lagrangians when starting from the scalar ones. Actually, at least all the terms that are functions of $A_\mu^\a$ and $S_{\mu\nu}^\a$ can be obtained from the scalar sector, which makes a very strong link between both theories. For that purpose, it is sufficient to consider the scalar multi-Galileon Lagrangian formulations given e.g. in Eqs.~\eqref{EqMultiGalVish} and~\eqref{EqExtGal}, and to do the replacement\footnote{Assuming that we consider Lagrangian forms which contain only first- and second-order derivatives of the scalar field, and no fields without any derivatives.} $\partial_\mu \pi^\a \rightarrow A_\mu^\a$. The second-order derivative of scalars can be promoted to $S_{\mu_i}{}^{\nu_i\a}$ or $G_{\mu_i}{}^{\nu_i\a}=\partial_{\mu_i} A^{\nu_i\a} - \partial^{\nu_i }A_{\mu_i}^\a$ terms, giving different kinds of terms. This procedure produces viable Lagrangians in the vector sector\footnote{It was shown in Ref.~\cite{Allys:2016jaq} that all Lagrangians built from contractions of one or two Levi-Civita tensors with vector Galileons and its first derivative propagate at most three degrees of freedom, as desired. This result, discussed in the single Galileon case, can be immediately extended to the multi-Galileon case.}.

The important point is that Lagrangians which are equivalent up to a total derivative in the scalar sector can not be related anymore when doing the replacement $\partial_\mu \pi^\a \rightarrow A_\mu^\a$. Indeed, the divergence of the associated currents could produce terms in $\partial_{\mu_i}\partial^{\nu_j}A_{\mu_k}^\a = (1/2)\partial^{\nu_j}G_{\mu_j\mu_k}^\a$, which vanish in the scalar sector. Several Lagrangians in the vector sector can then be obtained when starting from alternative equivalent formulations of a given dynamic in the scalar sector before promoting scalars to vectors. Such an examination was performed in Ref.~\cite{Allys:2016kbq}, where all the equivalent formulations of the pure scalar sector were detailed at a given order, and a comparison with the vector sector was done. Then, any examination of a vector multi-Galileon model should be performed in parallel to an examination of the associated scalar multi-Galileon model. It would be interesting to investigate this link in more detail in future works.


\section*{Acknowledgments} 

I wish to thank P.~Peter and Y.~Rodriguez for their support and advice, especially at the early stage of the project. I thank G.~Esposito-Farese for many valuable discussions and suggestions, and for a critical reading of the manuscript. I also thank C.~Deffayet, P.~Saffin and V.~Sivanesan for enlightening discussions, and T.~Marchand and J.~Van Dijk for their comments on my first draft.

\appendix

\section{Proof of property~\textbf{b}}
\label{AppProofB}

Before giving the proof of the property~\textbf{b}, let us recall some properties about the symmetry properties of a group of indices. To consider the symmetry properties under the exchange of an ensemble of variables, we will refer to the states with symmetry properties associated to Young diagrams, describing the different representations of the permutation group~\cite{Landau:1991wop}. These symmetrized state are such that the symmetry cannot be higher, which means that applying a symmetrization or anti-symmetrization on them gives either zero or a linear combination of states whose symmetry properties are related to other Young diagrams. One has to pay attention to the fact that symmetrizing or antisymmetrizing on two indices can change the symmetry properties of both these indices. For example, starting from a tensor that is symmetric under the exchange $\a\leftrightarrow\b$, and symmetrizing the indices $\b$ and $\c$, the result can be not symmetric anymore under the exchange $\a\leftrightarrow\b$. On the other hand, a (anti)symmetry on more than two indices implies a complete (anti)symmetry under exchange of the group of indices. For example, if there is a symmetry under the exchanges $\a\leftrightarrow\b$ and $\b\leftrightarrow\c$, then there is a complete symmetry for the $(\a,\b,\c)$ group of indices.

Let us first prove the property~\textbf{b} in the case of the $\mathcal{L}^{\text{ext}}_{_I}$, still using the notation of Eq.~\eqref{EqLExt1Cont}, i.e. considering a Lagrangian with an arbitrary $A^0_{[\a\b][\c\d]\e}$ prefactor. We will proceed in two steps, introducing at each step a current allowing us to improve the symmetry properties of the Lagrangian we started with. In a first step, we remove the contribution to $A^0_{[\a\b][\c\d]\e}$ that is totally antisymmetric for the group of indices $(\a,\b,\c,\d)$, which is a total derivative. To see it, one can take into account that the equations of motion will be linear combination of terms of the form $\delta^{\mu_1 \mu_2 \mu_3}_{\nu_1\nu_2\nu_3} \partial_{\mu_1}\pi^\al \partial^{\nu_1}\pi^\be\partial_{\mu_2}\partial^{\nu_2} \pi^{\color{red}\gamma\color{black}}\partial_{\mu_3}\partial^{\nu_3}\pi^{\color{red}\delta\color{black}}$ with internal indices contracted with four of the indices of $A_{[\a\b\c\d]\e}$. However, due to the fact that each term in the equations of motion is symmetric under the exchanges $\al\leftrightarrow\be$ and $\color{red}\gamma\color{black}\leftrightarrow\color{red}\delta\color{black}$, no non-vanishing contractions can be done with such an antisymmetric prefactor, and the equations of motion identically vanish. We call the new prefactor obtained this way $A^1_{[\a\b][\c\d]\e}$.

In fact, we removed in this step all the components of $A^0_{[\a\b][\c\d]\e}$ which are antisymmetric under the exchange of one index of $(\a,\b)$ and one index of $(\c,\d)$. To see it, let us assume for example that there is an antisymmetry on $\a\leftrightarrow\c$. Then, as there is a symmetry under the exchange $(\a,\b)\leftrightarrow(\c,\d)$, there is also an antisymmetry on $\b\leftrightarrow\d$. In addition, permuting e.g. $\a$ and $\c$ in the Lagrangian, and using the symmetry properties of $\delta^{\mu_1 \mu_2 \mu_3}_{\nu_1\nu_2\nu_3}$, one can show that the symmetric configurations in $\c\leftrightarrow\b$ and $\a\leftrightarrow\d$ identically vanish, which finally implies that there is a complete antisymmetry for the $(\a,\b,\c,\d)$ group of indices. This result is due to the strong symmetry conditions imposed by the structure of the extended multiGalileon Lagrangians, and will be useful in the following. It is then possible to symmetrize two pairs of indices between the $(\a,\b)$ and $(\c,\d)$ groups. Without loss of generality, we can consider that the pairs are $\a\leftrightarrow\c$ and $\b\leftrightarrow\d$, absorbing some minus signs in $A^1_{[\a\b][\c\d]\e}$ if necessary.

Let us now focus on the $(\a,\b,\e)$ group of indices. Using the property~\textbf{a}, it is possible to remove the contribution totally antisymmetric in $\a$, $\b$ and $\e$, which is a total derivative. The associated conserved current is obtained by removing $\partial^{\nu_3}$ from $\mathcal{L}^{\text{ext}}_{_I}$, and reads
\begin{equation}
\displaystyle{
J_{\nu_3}^{1} =\frac13 \delta^{\mu_1 \mu_2 \mu_3}_{\nu_1\nu_2\nu_3} \partial_{\mu_1}\pi^\a  
\partial_{\mu_2}\pi^\b\partial^{\nu_1}\pi^\c\partial^{\nu_2}\pi^\d \partial_{\mu_3}\pi^\e A^1_{[\a\b][\c\d]\e} = \frac13 \delta^{\mu_1 \mu_2 \mu_3}_{\nu_1\nu_2\nu_3} \partial_{\mu_1}\pi^{[\a}  
\partial_{\mu_2}\pi^\b\partial_{\mu_3}\pi^{\e]}\partial^{\nu_1}\pi^\c\partial^{\nu_2}\pi^\d  A^1_{[\a\b][\c\d]\e}.}
\end{equation}
The initial symmetry of the group of indices $(\a,\b,\e)$ being described by a linear combination of Young diagrams, and the totally antisymmetric one having been removed, each other terms are symmetric on at least two indices. This pair of indices cannot be $\a$ and $\b$, and we can always consider it as $\a$ and $\e$ (permuting $\partial_{\mu_1}$ and $\partial_{\mu_2}$ when necessary). In addition, the current $J_{\nu_1}^{1} $ does not contain a part which is antisymmetric on the group of indices $(\a,\b,\c,\d)$. Indeed, such a part would be in fact antisymmetric for the group of indices $(\a,\b,\c,\d,\e)$, and the only way to obtain it by antisymmetrizing on $\a\leftrightarrow\e$ (which entirely described the antisymmetrization done here due to the forced antisymmetry in $\a\leftrightarrow\b$) would be to start from a configuration already antisymmetric in $(\a,\b,\c,\d)$, which is excluded. In other words, and starting from a configuration which contains two pairs of indices that are symmetric by exchange, e.g. $\a\leftrightarrow\c$ and $\b\leftrightarrow\d$, it is not possible to obtain a completely antisymmetric configuration by antisymmetrizing some indices. 

The new prefactor obtained at this step is called $A^2_{[\a\b][\c\d]\e}$. This prefactor is symmetric by exchange under $\a\leftrightarrow\e$. As $A^2_{[\a\b][\c\d]\e}$ is a sum of two terms which do not contain any antisymmetric part in $(\a,\b,\c,\d)$, it can also be put in a form that is symmetric by exchange under $\a\leftrightarrow\c$ and $\b\leftrightarrow\d$ as explained previously. The only possibility for a term that is symmetric under the exchanges $\a\leftrightarrow\c$ and $\a\leftrightarrow\e$ is finally to be completely symmetric under exchange of $(\a,\c,\e)$ indices, which proves property \textbf{b}. It is indeed not possible to symmetrize on more indices, since this antisymmetrization would involve indices which are already antisymmetric by exchange. The demonstration of property~\textbf{b} is similar for the other Lagrangian terms. Note that this proof also allows us to obtain the additional symmetries of the Lagrangian. For $\mathcal{L}^{\text{ext}}_{_I}$ and $\mathcal{L}^{\text{ext}}_{_{II}}$, we can impose a symmetry under the exchange $\b\leftrightarrow\d$, and for $\mathcal{L}^{\text{ext}}_{_{III}}$ a symmetry under the exchanges $\b\leftrightarrow\e$ and $\c\leftrightarrow\f$ [using the notations of Eq.~\eqref{EqExtGal}].

Another proof of property~\textbf{b} can be done using Young diagrams to describe the symmetry properties under exchange of internal indices. Indeed, the symmetry under permutations of indices of the Lagrangians with a nontrivial dynamics can be associated to a Young diagram, or at least a linear combination of them. However, some symmetries are "forced", due to the presence of the $\delta^{\mu_1\cdots}_{\nu_1\cdots}$ term. These "forced" symmetries impose some subblocks of the Young diagrams describing the complete symmetry properties of the Lagrangians. Then, considering the possible complete diagrams formed from these subblocks, and taking into account the property~\textbf{a} as well as the form of the equations of motion, one recovers the result of property~\textbf{b}.

Let us apply it to $\mathcal{L}^{\text{ext}}_{_I}$, using the notation of Eq.~\eqref{EqLExt1Cont}, i.e. studying the symmetry properties of the $A_{[\a\b][\c\d]\e}$ prefactor. Due to the symmetrization of this tensor under the exchanges $\a\leftrightarrow\b$ and $\c\leftrightarrow\d$, the possible Young diagrams describing $A_{[\a\b][\c\d]\e}$ contain the following blocks:

\vspace{0.1cm}
\hfill
\begin{tabular}{|c|}
\hline
$\a$ \\
\hline
$\b$ \\
\hline
\end{tabular}
\hspace{2cm}
\begin{tabular}{|c|}
\hline
$\c$ \\
\hline
$\d$ \\
\hline
\end{tabular}
\hspace{2cm}
\begin{tabular}{|c|}
\hline
$\e$ \\
\hline
\end{tabular}
\hfill ~
\newline

Taking into account property~\textbf{a}, the Young diagrams describing $A_{[\a\b][\c\d]\e}$ cannot have the index $\e$ in the same column as any other indices. Thus, the only possible Young diagrams are 

\vspace{0.1cm}
\hfill
\begin{tabular}{c}
\begin{Young}
     $\a$  &  $\c$ &  $\e$ \cr
     $\b$  &  $\d$  \cr
\end{Young}
\end{tabular}
\hspace{2cm}
\begin{tabular}{c}
\begin{Young}
     $\a$  &  $\e$ \cr
     $\b$ \cr
     $\c$ \cr
     $\d$ \cr
\end{Young}
\end{tabular}
\hfill ~
\newline

The first configuration, which can be symmetrized on $(\a,\c,\e)$, is exactly the configuration described by property~\textbf{b}. Note that this configuration can also be symmetrized on $(\b,\d)$, as discussed previously. The second configuration is antisymmetric on $(\a,\b,\c,\d)$, and gives identically vanishing equations of motions, as explained before: it is a total derivative and can be omitted. This proves property \textbf{b}. 

Similar reasoning can be applied for $\mathcal{L}^{\text{ext}}_{_{II}}$ and $\mathcal{L}^{\text{ext}}_{_{III}}$, yielding respectively the following diagrams:

\vspace{0.1cm}
\hfill
\begin{tabular}{c}
\begin{Young}
     $\a$  &  $\c$ &  $\e$ & $\f$ \cr
     $\b$  &  $\d$  \cr
\end{Young}
\end{tabular}
\hspace{2cm}
\begin{tabular}{c}
\begin{Young}
     $\a$  &  $\d$ &  $\g$ \cr
     $\b$  &  $\e$  \cr
     $\c$  &  $\f$  \cr
\end{Young}
\end{tabular}
\hfill ~
\newline

These Young diagrams also show the additional symmetry properties discussed before.

\section{Terms with Levi-Civita tensors in the SO($N$) fundamental representation}
\label{AppEpsSON}

We consider in this section the possibility to build Lagrangians in the fundamental representation of a SO($N$) symmetry group with a prefactor containing Levi-Civita tensors. 
Note that property~\textbf{a}, in addition to the symmetry properties of the $\mathcal{L}^{\text{ext}}_j$ Lagrangians,
implies that no such Levi-Civita tensors can be contracted with $\partial_{\mu_i}\partial^{\nu_i} \pi^\a$ terms. Indeed, the other indices could be contracted either with a similar second-order derivative term, and the Lagrangian would identically vanish due to symmetry considerations, either with a first-order derivative term, and the Lagrangian would be a total derivative thanks to property~\textbf{a}. In addition, it is not necessary to consider contractions of Levi-Civita tensors together, since these contractions could be written with Kronecker delta only.

Let us begin with the case of SO(3), with a three-index Levi-Civita tensor $\epsilon_{\a\b\c}$. The only Lagrangian that can be written from $\mathcal{L}^{\text{ext}}_{_I}$ is
\begin{equation}
\mathcal{L}^{\text{SO}(3)}_4 = \displaystyle{ \delta^{\mu_1 \mu_2 \mu_3}_{\nu_1\nu_2\nu_3} \partial_{\mu_1}\pi^\a  
\partial_{\mu_2}\pi^\b\partial^{\nu_1}\pi^\c\partial^{\nu_2}\pi^\d \partial_{\mu_3}\partial^{\nu_3}\pi_\d \epsilon_{\a\b\c}}.
\end{equation}
Following property~\textbf{d}, we expect it to be a total derivative. It can be shown using the current
\begin{equation}
J^{3}_{\nu_3} =  \delta^{\mu_1 \mu_2 \mu_3}_{\nu_1\nu_2\nu_3} \partial_{\mu_1}\pi^\a  
\partial_{\mu_2}\pi^\b\partial^{\nu_1}\pi^\c\partial^{\nu_2}\pi^\d \partial_{\mu_3}\pi_\d \epsilon_{\a\b\c},
\end{equation}
in addition to property~\textbf{a}. No Lagrangian can be built from $\mathcal{L}^{\text{ext}}_{_{II}}$, since two Levi-Civita tensors would be involved, and thus there would be contractions between these tensors and $\partial_{\mu_i}\partial^{\nu_i} \pi^\a$ terms. With $\mathcal{L}^{\text{ext}}_{_{III}}$, only one term can be built,
\begin{equation}
\displaystyle{
\mathcal{L}^{\text{SO}(3)}_4 = \delta^{\mu_1 \mu_2 \mu_3 \mu_4}_{\nu_1\nu_2\nu_3\nu_4} \partial_{\mu_1}\pi^\a  \partial_{\mu_2}\pi^\b\partial_{\mu_3}\pi^\d\partial^{\nu_1}\pi^\c\partial^{\nu_2}\pi_\d\partial^{\nu_3}\pi^\e \partial_{\mu_4}\partial^{\nu_4}\pi_\e\epsilon_{\a\b\c}.
}
\end{equation}
However, we can show following property~\textbf{d} that it is a total derivative. The associated current is
\begin{equation}
\displaystyle{J^{4}_{\nu_4} = \delta^{\mu_1 \mu_2 \mu_3 \mu_4}_{\nu_1\nu_2\nu_3\nu_4} \partial_{\mu_1}\pi^\a  \partial_{\mu_2}\pi^\b\partial_{\mu_3}\pi^\d\partial^{\nu_1}\pi^\c\partial^{\nu_2}\pi_\d\partial^{\nu_3}\pi^\e \partial_{\mu_4}\pi_\e\epsilon_{\a\b\c},}
\end{equation}
which gives $\partial^{\nu_4} J^{4}_{\nu_4} = 2\mathcal{L}^{\text{SO}(N)}_4$.

Considering the case of SO(4), the only additional possible contraction is with $\mathcal{L}^{\text{ext}}_{_{II}}$, giving
\begin{equation}
\displaystyle{\mathcal{L}^{\text{SO}(4)}_5 =  \delta^{\mu_1 \mu_2 \mu_3 \mu_4}_{\nu_1\nu_2\nu_3\nu_4} \partial_{\mu_1}\pi^\a  \partial_{\mu_2}\pi^\b\partial^{\nu_1}\pi^\c\partial^{\nu_2}\pi^\d \partial_{\mu_3}\partial^{\nu_3}\pi^\e\partial_{\mu_4}\partial^{\nu_4}\pi_\e \epsilon_{\a\b\c\d}.}
\end{equation}
Following property~\textbf{d}, we can show that it is a total derivative, using the current
\begin{equation}
\displaystyle{J_{5}^{\mu_3} =  \delta^{\mu_1 \mu_2 \mu_3}_{\nu_1\nu_2\nu_3} \partial_{\mu_1}\pi^\a  
\partial_{\mu_2}\pi^\b\partial^{\nu_1}\pi^\c\partial^{\nu_2}\pi^\d  \partial^{\nu_3}\pi^\e\partial_{\mu_4}\partial^{\nu_4}\pi_\e \epsilon_{\a\b\c\d},}
\end{equation}
in addition to property~\textbf{a}.

For SO(5), the only possible contraction is with $\mathcal{L}^{\text{ext}}_{_{III}}$, yielding 
\begin{equation}
\mathcal{L}^{\text{SO}(5)}_6 = \delta^{\mu_1 \mu_2 \mu_3 \mu_4}_{\nu_1\nu_2\nu_3\nu_4} \partial_{\mu_1}\pi^\a  \partial_{\mu_2}\pi^\b\partial_{\mu_3}\pi^\c\partial^{\nu_1}\pi^\d\partial^{\nu_2}\pi^\e\partial^{\nu_3}\pi^\f \partial_{\mu_4}\partial^{\nu_4}\pi_\f \epsilon_{\a\b\c\d\e}.
\end{equation}
One more time, following property~\textbf{d}, we expect this Lagrangian to be a total derivative. It is shown by introducing the current 
\begin{equation}
\displaystyle{J_{6,\nu_4}=  \delta^{\mu_1 \mu_2 \mu_3 \mu_4}_{\nu_1\nu_2\nu_3\nu_4} \partial_{\mu_1}\pi^\a  \partial_{\mu_2}\pi^\b\partial_{\mu_3}\pi^\c\partial^{\nu_1}\pi^\d\partial^{\nu_2}\pi^\e\partial^{\nu_3}\pi^\f \partial_{\mu_4} \pi_\f \epsilon_{\a\b\c\d\e},}
\end{equation}
in addition to property~\textbf{a}.

\section{Exhaustive examination of $\mathcal{L}^{\text{ext}}_{_I}$ in the SU($N$) adjoint representation}
\label{AppPartSUN}
\subsection{Introduction}

We consider in this section all the possible independent Lagrangians which can be built from $\mathcal{L}^{\text{ext}}_{_I}$ in the case of multi-Galileons in the adjoint representation of a SU($N$) symmetry group, without using property~\textbf{b}. For this purpose, it is necessary to produce a basis of independent prefactors built from contractions of the primitive invariants only. The properties of the primitive invariants given in Sec.~\ref{PartIntroSUN} are thus very useful. One can also note that it is not necessary to consider too many contractions of indices. For example, the authors of Ref.~\cite{Dittner:1972hm} showed that the rank-seven or higher contractions of primitive invariants can be described with the contractions of lower ranks in the SU(3) case.  

These prefactors can be obtained from an explicit construction of the product representations. We give here an example for the singlet built from four adjoint representations of a SU(3) symmetry group. It is possible to build eight such singlets \cite{Slansky:1981yr,Feger:2012bs}. They can be identified through the product 
\begin{equation}
\boldsymbol{8} \times \boldsymbol{8}  = \boldsymbol{1} + \boldsymbol{8_s}+\boldsymbol{8_a} + \boldsymbol{10_a}+\boldsymbol{\overline{10}_a} + \boldsymbol{27_s},
\end{equation}
where the subscript $a$ or $s$ mean that the representations are symmetric or antisymmetric under the exchange of the two initial adjoint representations. Denoting for example the two initial adjoint representations as $\phi^\a$ and $\psi^\b$, the $\boldsymbol{8_s}$ representation is described by $d^{\a\b\c} \phi_\b \psi_\c$, and the $\boldsymbol{27_s}$ representation by $S^{\a\b}=\phi^{(\a}\psi^{\b)} + c.t.$, with the last term denoting counterterms such that $S_\a^\a=0$ and $d^{\a\b\c}S_{\b\c}=0$. This product allows us to identify the singlets from the product of four adjoint representations as those which appear in the product $\left(\boldsymbol{8} \times \boldsymbol{8}\right) \times \left(\boldsymbol{8} \times \boldsymbol{8}\right)$ as products of conjugate representations, since it is the only way to build a singlet from a product of two representations\footnote{We remind that the order of appearance of the different fields in this construction is not important. It is due to the fact that (denoting with $\boldsymbol{R_j}$ a given $j$ representation) if $\boldsymbol{R_1}\times\boldsymbol{R_2}$ contains $\boldsymbol{R_i}$, then $\boldsymbol{R_1}\times\boldsymbol{\bar R_i}$ contains $\boldsymbol{\bar R_2}$, etc.~\cite{Slansky:1981yr}}.

This method, even if exhaustive, becomes quite involved when high-dimensional representations appear in the products. In addition, it is not necessary to express the Lagrangians in terms of the irreducible representations which appear in the intermediate products of fields. Another equivalent method consists in listing all the independent prefactors built from primitive invariants at each order. For example, the only possibility with two fields is $\delta_{\a\b}$, and the two possibilities with three fields are $f_{\a\b\c}$ and $d_{\a\b\c}$. This number grows rapidly with the number of fields. The number of independent prefactors at each rank can be easily computed from group-theoretical calculations~\cite{Slansky:1981yr,Feger:2012bs}. The number of such combinations at different orders and for different SU($N$) groups are given in the following table.

\begin{center}
\begin{tabular}{|l|p{0.75cm}|p{0.75cm}|p{0.75cm}|p{0.75cm}|p{0.75cm}|p{0.75cm}|p{0.75cm}|}
\hline
$\#$ adjoint rep. & 1 & 2 & 3 & 4 & 5 & 6 & 7\\
\hline
$\#$ SU($2$) singlets & 0  & 1 & 1 & 3 & 6 & 15 & 36 \\
\hline
$\#$ SU($3$) singlets & 0  & 1 & 2 & 8 & 32 & 145 & 702 \\
\hline
$\#$ SU($4$) singlets & 0  & 1 & 2 & 9 & 43 & 245 & 1557 \\
\hline
$\#$ SU($5$) singlets & 0  & 1 & 2 & 9 & 44 & 264 & 1824 \\
\hline
\end{tabular}
\end{center}

We will use a third method in this paper. We saw in Sec.~\ref{PartFundRep} that the properties of the extended multi-Galileon Lagrangians are quite restrictive, particularly the symmetry properties. Thus, contractions with only a small number of primitive invariant prefactors will allow nontrivial dynamics at first glance. We will then proceed by considering the possible contractions for each kind of primitive invariant prefactor. After having obtained the nontrivial Lagrangians, we will finally verify their independence taking into account the relations between the primitive invariants contractions.

\subsection{Terms without primitive invariants contracted together}

The first possibility is to contract $\mathcal{L}^{\text{ext}}_{_I}$ with one Kronecker delta and one structure constant. However, this term has already been investigated in the case of the fundamental 
representation of a SO($3$) symmetry group, since the only property of antisymmetry of $\epsilon_{\a\b\c}$ has been used. One nontrivial dynamic is possible, described e.g. by the Lagrangian given in Eq.~\eqref{EqLagFinSON} whose equation of motion is given in Eq.~\eqref{EqEOMSON}.

The other possibility is to contract $\mathcal{L}^{\text{ext}}_{_I}$ with one Kronecker delta and one symmetric $d_{\a\b\c}$ invariant. Only one such Lagrangian is possible, i.e. 
\begin{equation}
\displaystyle{\mathcal{L}^{\text{SU}(N)}_1 =   \delta^{\mu_1 \mu_2 \mu_3}_{\nu_1\nu_2\nu_3} \partial_{\mu_1}\pi^\a \partial_{\mu_2}\pi^\d\partial^{\nu_1}\pi^\b\partial^{\nu_2}\pi_\d \partial_{\mu_3}\partial^{\nu_3}\pi^\c d_{\a\b\c}.}
\end{equation}
Property~\textbf{d} cannot be applied here, and this Lagrangian is not a total derivative. Indeed, the currents that can be formed by removing $\partial_{\mu_3}$ or $\partial^{\nu_3}$ from this Lagrangian trivially vanish. The equation of motion thus gives
\begin{multline}
\displaystyle{EOM^{\text{SU}(N)}_1 = 
\delta^{\mu_1 \mu_2 \mu_3}_{\nu_1\nu_2\nu_3} \left[
4 \partial_{\mu_1}\pi^\b \partial^{\nu_1}\pi^\d \partial_{\mu_2}\partial^{\nu_2}\pi^\c \partial_{\mu_3}\partial^{\nu_3}\pi_\d d^{\al}{}_{\b\c}
-3 \partial_{\mu_1}\pi^\d \partial^{\nu_1}\pi_\d \partial_{\mu_2}\partial^{\nu_2}\pi^\b \partial_{\mu_3}\partial^{\nu_3}\pi^\c d^{\al}{}_{\b\c}
\right.} \\ \displaystyle{\left.
- \partial_{\mu_1}\pi^\b \partial^{\nu_1}\pi^\c\partial_{\mu_2}\partial^{\nu_2}\pi^\d \partial_{\mu_3}\partial^{\nu_3}\pi_\d d^{\al}{}_{\b\c}
+2 \partial_{\mu_1}\pi^\a \partial^{\nu_1}\pi^\al\partial_{\mu_2}\partial^{\nu_2}\pi^\b \partial_{\mu_3}\partial^{\nu_3}\pi^\c d_{\a\b\c}
\right.} \\ \displaystyle{\left.
-2 \partial_{\mu_1}\pi^\a \partial^{\nu_1}\pi^\b\partial_{\mu_2}\partial^{\nu_2}\pi^\al \partial_{\mu_3}\partial^{\nu_3}\pi^\c d_{\a\b\c},
\right]}
\label{AppEqEOMSUN}
\end{multline}
where $\al$ is the free group index of the equation of motion, since this equation of motion is in the adjoint representation of the SU($N$) symmetry group.
This equation of motion is a total derivative, as expected. Indeed, introducing the currents
\begin{equation}
\displaystyle{\left\{
\begin{array}{l}
J_1^{\mu_2} = \delta^{\mu_1 \mu_2 \mu_3}_{\nu_1\nu_2\nu_3} \partial_{\mu_1}\pi^\a \partial^{\nu_1}\pi^\al\partial^{\nu_2}\pi^\b \partial_{\mu_3}\partial^{\nu_3}\pi^\c d_{\a\b\c},\\
J_2^{\mu_2} = \delta^{\mu_1 \mu_2 \mu_3}_{\nu_1\nu_2\nu_3} \partial_{\mu_1}\pi^\b \partial^{\nu_1}\pi^\d\partial^{\nu_2}\pi^\c \partial_{\mu_3}\partial^{\nu_3}\pi_\d d^{\al}{}_{\b\c},\\
J_3^{\mu_2} = \delta^{\mu_1 \mu_2 \mu_3}_{\nu_1\nu_2\nu_3}\partial_{\mu_1}\pi_\d \partial^{\nu_1}\pi^\b\partial^{\nu_2}\pi^\d \partial_{\mu_3}\partial^{\nu_3}\pi^\c d^{\al}{}_{\b\c},
\end{array}
\right.}
\end{equation}
one can verify that 
\begin{equation}
EOM^{\text{SU}(N)}_1 = \partial_{\mu_2} \left[2 J_1^{\mu_2} + J_2^{\mu_2} + 3 J_3^{\mu_2} \right].
\end{equation}

\subsection{Terms with primitive invariants contracted together}

As discussed before, it is not necessary to consider the contractions between $\delta_{\a\b}$ and the other primitive invariants, since they only raise or lower the indices. We thus focus on the contractions of $f_{\a\b\c}$ and $d_{\a\b\c}$. As $\mathcal{L}^{\text{ext}}_{_I}$ contains five fields, it is sufficient to consider only the rank-five contractions, i.e. with five indices not contracted together. Indeed, considering the rank-four contractions, it would then be necessary to form a singlet from the single remaining field, which is not possible. To describe the rank-five contractions of the primitive invariants, it is sufficient to consider the contractions of only three $f$ or $d$. We will then consider them successively. The contractions of more primitive invariants will only reduce to those already considered thanks to the different structure properties of the group.

Concerning the contractions with a prefactor of the form $f_{\be \a\b}f_{\ga \c\d}f^{\be\ga}{}_{\e}$, no contractions are possible due to symmetry considerations. Concerning the prefactor of the form $f_{\be \a\b}d_{\ga \c\d}d^{\be\ga}{}_{\e}$, two Lagrangians can be written, yielding
\begin{equation}
\displaystyle{\begin{array}{l}
\mathcal{L}^{\text{SU}(N)}_2 =   \delta^{\mu_1 \mu_2 \mu_3}_{\nu_1\nu_2\nu_3} \partial_{\mu_1}\pi^\a \partial_{\mu_2}\pi^\b\partial^{\nu_1}\pi^\c\partial^{\nu_2}\pi^\e \partial_{\mu_3}\partial^{\nu_3}\pi^\d f_{\be \a\b}d_{\ga \c\d}d^{\be\ga}{}_{\e},\\
\mathcal{L}^{\text{SU}(N)}_3 =   \delta^{\mu_1 \mu_2 \mu_3}_{\nu_1\nu_2\nu_3} \partial_{\mu_1}\pi^\a \partial_{\mu_2}\pi^\c\partial^{\nu_1}\pi^\b\partial^{\nu_2}\pi^\e \partial_{\mu_3}\partial^{\nu_3}\pi^\d f_{\be \a\b}d_{\ga \c\d}d^{\be\ga}{}_{\e}.
\end{array}}
\end{equation}
However, building currents by removing $\partial^{\nu_3}$ from $\mathcal{L}^{\text{SU}(N)}_2$ and $\partial_{\mu_3}$ from $\mathcal{L}^{\text{SU}(N)}_3$, and using property~\textbf{a} as well as symmetry properties, one can show that both Lagrangians are total derivatives.
The two Lagrangians built from the contractions with $f_{\be \a\b}f_{\ga \c\d}d^{\be\ga}{}_{\e}$, as well as both Lagrangians built with $f_{\be \a\b}d_{\ga \c\d}d^{\be\ga}{}_{\e}$, are also total derivatives for similar reasons. 

Let us turn to the contractions with a prefactor in $d_{\be \a\b}d_{\ga \c\d}f^{\be\ga}{}_{\e}$. This prefactor allows us to build only one Lagrangian, 
\begin{equation}
\displaystyle{
\mathcal{L}^{\text{SU}(N)}_4 =   \delta^{\mu_1 \mu_2 \mu_3}_{\nu_1\nu_2\nu_3} \partial_{\mu_1}\pi^\a \partial_{\mu_2}\pi^\c\partial^{\nu_1}\pi^\b\partial^{\nu_2}\pi^\e \partial_{\mu_3}\partial^{\nu_3}\pi^\d d_{\be \a\b}d_{\ga \c\d}f^{\be\ga}{}_{\e}.
}
\end{equation}
This Lagrangian cannot be written as a total derivative, since the currents obtained by removing $\partial_{\mu_3}$ or $\partial^{\nu_3}$ vanish thanks to symmetry considerations. However, it is possible to link this Lagrangian with other ones thanks to the structure properties of SU($N$). The primitive invariants verify the following relations (see e.g. Ref.~\cite{Fuchs:1997jv})
\begin{equation}
\label{EqPropStructSUN}
\displaystyle{
f_{\a\d\be}d^{\be}{}_{\b\c} + f_{\b\d\be}d^{\be}{}_{\c\a} +f_{\c\d\be}d^{\be}{}_{\a\b} =0.
}
\end{equation}
This relation implies that 
\begin{equation}
\mathcal{L}^{\text{SU}(N)}_4 = \mathcal{L}^{\text{SU}(N)}_2 + \mathcal{L}^{\text{SU}(N)}_3,
\end{equation}
and thus that $\mathcal{L}^{\text{SU}(N)}_4$ is a total derivative. Note that this result cannot be seen directly from the equation of motion of $\mathcal{L}^{\text{SU}(N)}_4$ without using Eq.~\eqref{EqPropStructSUN}. It is due to the fact that we used a basis of primitive invariant prefactors which is convenient, but with terms which could be not linearly independent.

Finally, one can consider the possible contraction with a $d_{\be \a\b} d_{\ga \c\d} d^{\be\ga}{}_{\e}$ prefactor. It is possible to build two such Lagrangians, i.e. 
\begin{equation}
\displaystyle{\begin{array}{l}
\mathcal{L}^{\text{SU}(N)}_5 =   \delta^{\mu_1 \mu_2 \mu_3}_{\nu_1\nu_2\nu_3} \partial_{\mu_1}\pi^\a \partial_{\mu_2}\pi^\c\partial^{\nu_1}\pi^\b\partial^{\nu_2}\pi^\d \partial_{\mu_3}\partial^{\nu_3}\pi^\e d_{\be \a\b} d_{\ga \c\d} d^{\be\ga}{}_{\e},\\
\mathcal{L}^{\text{SU}(N)}_6 =   \delta^{\mu_1 \mu_2 \mu_3}_{\nu_1\nu_2\nu_3} \partial_{\mu_1}\pi^\a \partial_{\mu_2}\pi^\c\partial^{\nu_1}\pi^\b\partial^{\nu_2}\pi^\e \partial_{\mu_3}\partial^{\nu_3}\pi^\d d_{\be \a\b} d_{\ga \c\d} d^{\be\ga}{}_{\e}.
\end{array}}
\end{equation}
As expected from property~\textbf{e}, they are related by a total derivative. Indeed, the current:
\begin{equation}
\displaystyle{
J^{\mu_3} = \delta^{\mu_1 \mu_2 \mu_3}_{\nu_1\nu_2\nu_3}  \partial_{\mu_1}\pi^\a \partial_{\mu_2}\pi^\c\partial^{\nu_1}\pi^\b\partial^{\nu_2}\pi^\d \partial^{\nu_3}\pi^\e d_{\be \a\b} d_{\ga \c\d} d^{\be\ga}{}_{\e},
}
\end{equation}
gives
\begin{equation}
\displaystyle{
\partial_{\mu_3}J^{\mu_3} = \mathcal{L}^{\text{SU}(N)}_5 - 2 \mathcal{L}^{\text{SU}(N)}_6.
}
\end{equation}
Then, it is also possible to use relations between the primitive invariants of SU($N$), especially \cite{Fuchs:1997jv}
\begin{equation}
\label{EqPropStructSUN2}
\displaystyle{
f_{\a\b\be}f_{\c\d}{}^{\be}= \frac2N\left(\delta_{\a\c}\delta_{\b\d} - \delta_{\b\c}\delta_{\a\d} \right) + d_{\a\c\be}d_{\b\d}{}^{\be} - d_{\a\d\be}d_{\b\c}{}^{\be},
}
\end{equation}
which allows us to write down 
\begin{equation}
\displaystyle{
\mathcal{L}^{\text{SU}(N)}_6 - \mathcal{L}^{\text{SU}(N)}_5 = \frac2N\mathcal{L}^{\text{SU}(N)}_1 + \text{total derivative},
}
\end{equation}
with the total derivative being obtained thanks to property~\textbf{a}. It implies that both $\mathcal{L}^{\text{SU}(N)}_5$ and $\mathcal{L}^{\text{SU}(N)}_6$ involve the same dynamics as $\mathcal{L}^{\text{SU}(N)}_1$. 

This finally shows that only one nontrivial dynamics allowed at the order of $\mathcal{L}^{\text{ext}}_{_I}$ is described by the Lagrangian $\mathcal{L}^{\text{SU}(N)}_1$. It is exactly what is predicted from the construction of Sec.~\ref{PartSUNLag} with a third-rank symmetric tensor as implied by property~\textbf{b}.


%


\begin{thebibliography}{82}%
\makeatletter
\providecommand \@ifxundefined [1]{%
 \@ifx{#1\undefined}
}%
\providecommand \@ifnum [1]{%
 \ifnum #1\expandafter \@firstoftwo
 \else \expandafter \@secondoftwo
 \fi
}%
\providecommand \@ifx [1]{%
 \ifx #1\expandafter \@firstoftwo
 \else \expandafter \@secondoftwo
 \fi
}%
\providecommand \natexlab [1]{#1}%
\providecommand \enquote  [1]{``#1''}%
\providecommand \bibnamefont  [1]{#1}%
\providecommand \bibfnamefont [1]{#1}%
\providecommand \citenamefont [1]{#1}%
\providecommand \href@noop [0]{\@secondoftwo}%
\providecommand \href [0]{\begingroup \@sanitize@url \@href}%
\providecommand \@href[1]{\@@startlink{#1}\@@href}%
\providecommand \@@href[1]{\endgroup#1\@@endlink}%
\providecommand \@sanitize@url [0]{\catcode `\\12\catcode `\$12\catcode
  `\&12\catcode `\#12\catcode `\^12\catcode `\_12\catcode `\%12\relax}%
\providecommand \@@startlink[1]{}%
\providecommand \@@endlink[0]{}%
\providecommand \url  [0]{\begingroup\@sanitize@url \@url }%
\providecommand \@url [1]{\endgroup\@href {#1}{\urlprefix }}%
\providecommand \urlprefix  [0]{URL }%
\providecommand \Eprint [0]{\href }%
\providecommand \doibase [0]{http://dx.doi.org/}%
\providecommand \selectlanguage [0]{\@gobble}%
\providecommand \bibinfo  [0]{\@secondoftwo}%
\providecommand \bibfield  [0]{\@secondoftwo}%
\providecommand \translation [1]{[#1]}%
\providecommand \BibitemOpen [0]{}%
\providecommand \bibitemStop [0]{}%
\providecommand \bibitemNoStop [0]{.\EOS\space}%
\providecommand \EOS [0]{\spacefactor3000\relax}%
\providecommand \BibitemShut  [1]{\csname bibitem#1\endcsname}%
\let\auto@bib@innerbib\@empty
\bibitem [{\citenamefont {Nicolis}\ \emph {et~al.}(2009)\citenamefont
  {Nicolis}, \citenamefont {Rattazzi},\ and\ \citenamefont
  {Trincherini}}]{Nicolis:2008in}%
  \BibitemOpen
  \bibfield  {author} {\bibinfo {author} {\bibfnamefont {A.}~\bibnamefont
  {Nicolis}}, \bibinfo {author} {\bibfnamefont {R.}~\bibnamefont {Rattazzi}}, \
  and\ \bibinfo {author} {\bibfnamefont {E.}~\bibnamefont {Trincherini}},\
  }\href {\doibase 10.1103/PhysRevD.79.064036} {\bibfield  {journal} {\bibinfo
  {journal} {Phys.Rev.}\ }\textbf {\bibinfo {volume} {D79}},\ \bibinfo {pages}
  {064036} (\bibinfo {year} {2009})},\ \Eprint {http://arxiv.org/abs/0811.2197}
  {arXiv:0811.2197 [hep-th]} \BibitemShut {NoStop}%
\bibitem [{\citenamefont {Deffayet}\ \emph
  {et~al.}(2009{\natexlab{a}})\citenamefont {Deffayet}, \citenamefont
  {Esposito-Farese},\ and\ \citenamefont {Vikman}}]{Deffayet:2009wt}%
  \BibitemOpen
  \bibfield  {author} {\bibinfo {author} {\bibfnamefont {C.}~\bibnamefont
  {Deffayet}}, \bibinfo {author} {\bibfnamefont {G.}~\bibnamefont
  {Esposito-Farese}}, \ and\ \bibinfo {author} {\bibfnamefont {A.}~\bibnamefont
  {Vikman}},\ }\href {\doibase 10.1103/PhysRevD.79.084003} {\bibfield
  {journal} {\bibinfo  {journal} {Phys.Rev.}\ }\textbf {\bibinfo {volume}
  {D79}},\ \bibinfo {pages} {084003} (\bibinfo {year} {2009}{\natexlab{a}})},\
  \Eprint {http://arxiv.org/abs/0901.1314} {arXiv:0901.1314 [hep-th]}
  \BibitemShut {NoStop}%
\bibitem [{\citenamefont {Deffayet}\ \emph
  {et~al.}(2009{\natexlab{b}})\citenamefont {Deffayet}, \citenamefont {Deser},\
  and\ \citenamefont {Esposito-Farese}}]{Deffayet:2009mn}%
  \BibitemOpen
  \bibfield  {author} {\bibinfo {author} {\bibfnamefont {C.}~\bibnamefont
  {Deffayet}}, \bibinfo {author} {\bibfnamefont {S.}~\bibnamefont {Deser}}, \
  and\ \bibinfo {author} {\bibfnamefont {G.}~\bibnamefont {Esposito-Farese}},\
  }\href {\doibase 10.1103/PhysRevD.80.064015} {\bibfield  {journal} {\bibinfo
  {journal} {Phys.Rev.}\ }\textbf {\bibinfo {volume} {D80}},\ \bibinfo {pages}
  {064015} (\bibinfo {year} {2009}{\natexlab{b}})},\ \Eprint
  {http://arxiv.org/abs/0906.1967} {arXiv:0906.1967 [gr-qc]} \BibitemShut
  {NoStop}%
\bibitem [{\citenamefont {Deffayet}\ \emph {et~al.}(2011)\citenamefont
  {Deffayet}, \citenamefont {Gao}, \citenamefont {Steer},\ and\ \citenamefont
  {Zahariade}}]{Deffayet:2011gz}%
  \BibitemOpen
  \bibfield  {author} {\bibinfo {author} {\bibfnamefont {C.}~\bibnamefont
  {Deffayet}}, \bibinfo {author} {\bibfnamefont {X.}~\bibnamefont {Gao}},
  \bibinfo {author} {\bibfnamefont {D.}~\bibnamefont {Steer}}, \ and\ \bibinfo
  {author} {\bibfnamefont {G.}~\bibnamefont {Zahariade}},\ }\href {\doibase
  10.1103/PhysRevD.84.064039} {\bibfield  {journal} {\bibinfo  {journal}
  {Phys.Rev.}\ }\textbf {\bibinfo {volume} {D84}},\ \bibinfo {pages} {064039}
  (\bibinfo {year} {2011})},\ \Eprint {http://arxiv.org/abs/1103.3260}
  {arXiv:1103.3260 [hep-th]} \BibitemShut {NoStop}%
\bibitem [{\citenamefont {de~Rham}\ and\ \citenamefont
  {Heisenberg}(2011)}]{deRham:2011by}%
  \BibitemOpen
  \bibfield  {author} {\bibinfo {author} {\bibfnamefont {C.}~\bibnamefont
  {de~Rham}}\ and\ \bibinfo {author} {\bibfnamefont {L.}~\bibnamefont
  {Heisenberg}},\ }\href {\doibase 10.1103/PhysRevD.84.043503} {\bibfield
  {journal} {\bibinfo  {journal} {Phys. Rev.}\ }\textbf {\bibinfo {volume}
  {D84}},\ \bibinfo {pages} {043503} (\bibinfo {year} {2011})},\ \Eprint
  {http://arxiv.org/abs/1106.3312} {arXiv:1106.3312 [hep-th]} \BibitemShut
  {NoStop}%
\bibitem [{\citenamefont {Joyce}\ \emph {et~al.}(2015)\citenamefont {Joyce},
  \citenamefont {Jain}, \citenamefont {Khoury},\ and\ \citenamefont
  {Trodden}}]{Joyce:2014kja}%
  \BibitemOpen
  \bibfield  {author} {\bibinfo {author} {\bibfnamefont {A.}~\bibnamefont
  {Joyce}}, \bibinfo {author} {\bibfnamefont {B.}~\bibnamefont {Jain}},
  \bibinfo {author} {\bibfnamefont {J.}~\bibnamefont {Khoury}}, \ and\ \bibinfo
  {author} {\bibfnamefont {M.}~\bibnamefont {Trodden}},\ }\href {\doibase
  10.1016/j.physrep.2014.12.002} {\bibfield  {journal} {\bibinfo  {journal}
  {Phys. Rept.}\ }\textbf {\bibinfo {volume} {568}},\ \bibinfo {pages} {1}
  (\bibinfo {year} {2015})},\ \Eprint {http://arxiv.org/abs/1407.0059}
  {arXiv:1407.0059 [astro-ph.CO]} \BibitemShut {NoStop}%
\bibitem [{\citenamefont {Chow}\ and\ \citenamefont
  {Khoury}(2009)}]{Chow:2009fm}%
  \BibitemOpen
  \bibfield  {author} {\bibinfo {author} {\bibfnamefont {N.}~\bibnamefont
  {Chow}}\ and\ \bibinfo {author} {\bibfnamefont {J.}~\bibnamefont {Khoury}},\
  }\href {\doibase 10.1103/PhysRevD.80.024037} {\bibfield  {journal} {\bibinfo
  {journal} {Phys. Rev.}\ }\textbf {\bibinfo {volume} {D80}},\ \bibinfo {pages}
  {024037} (\bibinfo {year} {2009})},\ \Eprint {http://arxiv.org/abs/0905.1325}
  {arXiv:0905.1325 [hep-th]} \BibitemShut {NoStop}%
\bibitem [{\citenamefont {Silva}\ and\ \citenamefont
  {Koyama}(2009)}]{Silva:2009km}%
  \BibitemOpen
  \bibfield  {author} {\bibinfo {author} {\bibfnamefont {F.~P.}\ \bibnamefont
  {Silva}}\ and\ \bibinfo {author} {\bibfnamefont {K.}~\bibnamefont {Koyama}},\
  }\href {\doibase 10.1103/PhysRevD.80.121301} {\bibfield  {journal} {\bibinfo
  {journal} {Phys. Rev.}\ }\textbf {\bibinfo {volume} {D80}},\ \bibinfo {pages}
  {121301} (\bibinfo {year} {2009})},\ \Eprint {http://arxiv.org/abs/0909.4538}
  {arXiv:0909.4538 [astro-ph.CO]} \BibitemShut {NoStop}%
\bibitem [{\citenamefont {Deffayet}\ \emph
  {et~al.}(2010{\natexlab{a}})\citenamefont {Deffayet}, \citenamefont
  {Pujolas}, \citenamefont {Sawicki},\ and\ \citenamefont
  {Vikman}}]{Deffayet:2010qz}%
  \BibitemOpen
  \bibfield  {author} {\bibinfo {author} {\bibfnamefont {C.}~\bibnamefont
  {Deffayet}}, \bibinfo {author} {\bibfnamefont {O.}~\bibnamefont {Pujolas}},
  \bibinfo {author} {\bibfnamefont {I.}~\bibnamefont {Sawicki}}, \ and\
  \bibinfo {author} {\bibfnamefont {A.}~\bibnamefont {Vikman}},\ }\href
  {\doibase 10.1088/1475-7516/2010/10/026} {\bibfield  {journal} {\bibinfo
  {journal} {JCAP}\ }\textbf {\bibinfo {volume} {1010}},\ \bibinfo {pages}
  {026} (\bibinfo {year} {2010}{\natexlab{a}})},\ \Eprint
  {http://arxiv.org/abs/1008.0048} {arXiv:1008.0048 [hep-th]} \BibitemShut
  {NoStop}%
\bibitem [{\citenamefont {Kobayashi}(2010)}]{Kobayashi:2010wa}%
  \BibitemOpen
  \bibfield  {author} {\bibinfo {author} {\bibfnamefont {T.}~\bibnamefont
  {Kobayashi}},\ }\href {\doibase 10.1103/PhysRevD.81.103533} {\bibfield
  {journal} {\bibinfo  {journal} {Phys. Rev.}\ }\textbf {\bibinfo {volume}
  {D81}},\ \bibinfo {pages} {103533} (\bibinfo {year} {2010})},\ \Eprint
  {http://arxiv.org/abs/1003.3281} {arXiv:1003.3281 [astro-ph.CO]} \BibitemShut
  {NoStop}%
\bibitem [{\citenamefont {Gannouji}\ and\ \citenamefont
  {Sami}(2010)}]{Gannouji:2010au}%
  \BibitemOpen
  \bibfield  {author} {\bibinfo {author} {\bibfnamefont {R.}~\bibnamefont
  {Gannouji}}\ and\ \bibinfo {author} {\bibfnamefont {M.}~\bibnamefont
  {Sami}},\ }\href {\doibase 10.1103/PhysRevD.82.024011} {\bibfield  {journal}
  {\bibinfo  {journal} {Phys. Rev.}\ }\textbf {\bibinfo {volume} {D82}},\
  \bibinfo {pages} {024011} (\bibinfo {year} {2010})},\ \Eprint
  {http://arxiv.org/abs/1004.2808} {arXiv:1004.2808 [gr-qc]} \BibitemShut
  {NoStop}%
\bibitem [{\citenamefont {Tsujikawa}(2010)}]{Tsujikawa:2010zza}%
  \BibitemOpen
  \bibfield  {author} {\bibinfo {author} {\bibfnamefont {S.}~\bibnamefont
  {Tsujikawa}},\ }\href {\doibase 10.1007/978-3-642-10598-2_3} {\bibfield
  {journal} {\bibinfo  {journal} {Lect. Notes Phys.}\ }\textbf {\bibinfo
  {volume} {800}},\ \bibinfo {pages} {99} (\bibinfo {year} {2010})},\ \Eprint
  {http://arxiv.org/abs/1101.0191} {arXiv:1101.0191 [gr-qc]} \BibitemShut
  {NoStop}%
\bibitem [{\citenamefont {De~Felice}\ and\ \citenamefont
  {Tsujikawa}(2010)}]{DeFelice:2010pv}%
  \BibitemOpen
  \bibfield  {author} {\bibinfo {author} {\bibfnamefont {A.}~\bibnamefont
  {De~Felice}}\ and\ \bibinfo {author} {\bibfnamefont {S.}~\bibnamefont
  {Tsujikawa}},\ }\href {\doibase 10.1103/PhysRevLett.105.111301} {\bibfield
  {journal} {\bibinfo  {journal} {Phys. Rev. Lett.}\ }\textbf {\bibinfo
  {volume} {105}},\ \bibinfo {pages} {111301} (\bibinfo {year} {2010})},\
  \Eprint {http://arxiv.org/abs/1007.2700} {arXiv:1007.2700 [astro-ph.CO]}
  \BibitemShut {NoStop}%
\bibitem [{\citenamefont {Ali}\ \emph {et~al.}(2010)\citenamefont {Ali},
  \citenamefont {Gannouji},\ and\ \citenamefont {Sami}}]{Ali:2010gr}%
  \BibitemOpen
  \bibfield  {author} {\bibinfo {author} {\bibfnamefont {A.}~\bibnamefont
  {Ali}}, \bibinfo {author} {\bibfnamefont {R.}~\bibnamefont {Gannouji}}, \
  and\ \bibinfo {author} {\bibfnamefont {M.}~\bibnamefont {Sami}},\ }\href
  {\doibase 10.1103/PhysRevD.82.103015} {\bibfield  {journal} {\bibinfo
  {journal} {Phys. Rev.}\ }\textbf {\bibinfo {volume} {D82}},\ \bibinfo {pages}
  {103015} (\bibinfo {year} {2010})},\ \Eprint {http://arxiv.org/abs/1008.1588}
  {arXiv:1008.1588 [astro-ph.CO]} \BibitemShut {NoStop}%
\bibitem [{\citenamefont {De~Felice}\ and\ \citenamefont
  {Tsujikawa}(2011)}]{DeFelice:2010nf}%
  \BibitemOpen
  \bibfield  {author} {\bibinfo {author} {\bibfnamefont {A.}~\bibnamefont
  {De~Felice}}\ and\ \bibinfo {author} {\bibfnamefont {S.}~\bibnamefont
  {Tsujikawa}},\ }\href {\doibase 10.1103/PhysRevD.84.124029} {\bibfield
  {journal} {\bibinfo  {journal} {Phys. Rev.}\ }\textbf {\bibinfo {volume}
  {D84}},\ \bibinfo {pages} {124029} (\bibinfo {year} {2011})},\ \Eprint
  {http://arxiv.org/abs/1008.4236} {arXiv:1008.4236 [hep-th]} \BibitemShut
  {NoStop}%
\bibitem [{\citenamefont {Mota}\ \emph {et~al.}(2010)\citenamefont {Mota},
  \citenamefont {Sandstad},\ and\ \citenamefont {Zlosnik}}]{Mota:2010bs}%
  \BibitemOpen
  \bibfield  {author} {\bibinfo {author} {\bibfnamefont {D.~F.}\ \bibnamefont
  {Mota}}, \bibinfo {author} {\bibfnamefont {M.}~\bibnamefont {Sandstad}}, \
  and\ \bibinfo {author} {\bibfnamefont {T.}~\bibnamefont {Zlosnik}},\ }\href
  {\doibase 10.1007/JHEP12(2010)051} {\bibfield  {journal} {\bibinfo  {journal}
  {JHEP}\ }\textbf {\bibinfo {volume} {1012}},\ \bibinfo {pages} {051}
  (\bibinfo {year} {2010})},\ \Eprint {http://arxiv.org/abs/1009.6151}
  {arXiv:1009.6151 [astro-ph.CO]} \BibitemShut {NoStop}%
\bibitem [{\citenamefont {Nesseris}\ \emph {et~al.}(2010)\citenamefont
  {Nesseris}, \citenamefont {De~Felice},\ and\ \citenamefont
  {Tsujikawa}}]{Nesseris:2010pc}%
  \BibitemOpen
  \bibfield  {author} {\bibinfo {author} {\bibfnamefont {S.}~\bibnamefont
  {Nesseris}}, \bibinfo {author} {\bibfnamefont {A.}~\bibnamefont {De~Felice}},
  \ and\ \bibinfo {author} {\bibfnamefont {S.}~\bibnamefont {Tsujikawa}},\
  }\href {\doibase 10.1103/PhysRevD.82.124054} {\bibfield  {journal} {\bibinfo
  {journal} {Phys. Rev.}\ }\textbf {\bibinfo {volume} {D82}},\ \bibinfo {pages}
  {124054} (\bibinfo {year} {2010})},\ \Eprint {http://arxiv.org/abs/1010.0407}
  {arXiv:1010.0407 [astro-ph.CO]} \BibitemShut {NoStop}%
\bibitem [{\citenamefont {Easson}\ \emph {et~al.}(2011)\citenamefont {Easson},
  \citenamefont {Sawicki},\ and\ \citenamefont {Vikman}}]{Easson:2011zy}%
  \BibitemOpen
  \bibfield  {author} {\bibinfo {author} {\bibfnamefont {D.~A.}\ \bibnamefont
  {Easson}}, \bibinfo {author} {\bibfnamefont {I.}~\bibnamefont {Sawicki}}, \
  and\ \bibinfo {author} {\bibfnamefont {A.}~\bibnamefont {Vikman}},\ }\href
  {\doibase 10.1088/1475-7516/2011/11/021} {\bibfield  {journal} {\bibinfo
  {journal} {JCAP}\ }\textbf {\bibinfo {volume} {1111}},\ \bibinfo {pages}
  {021} (\bibinfo {year} {2011})},\ \Eprint {http://arxiv.org/abs/1109.1047}
  {arXiv:1109.1047 [hep-th]} \BibitemShut {NoStop}%
\bibitem [{\citenamefont {Gleyzes}\ \emph {et~al.}(2013)\citenamefont
  {Gleyzes}, \citenamefont {Langlois}, \citenamefont {Piazza},\ and\
  \citenamefont {Vernizzi}}]{Gleyzes:2013ooa}%
  \BibitemOpen
  \bibfield  {author} {\bibinfo {author} {\bibfnamefont {J.}~\bibnamefont
  {Gleyzes}}, \bibinfo {author} {\bibfnamefont {D.}~\bibnamefont {Langlois}},
  \bibinfo {author} {\bibfnamefont {F.}~\bibnamefont {Piazza}}, \ and\ \bibinfo
  {author} {\bibfnamefont {F.}~\bibnamefont {Vernizzi}},\ }\href {\doibase
  10.1088/1475-7516/2013/08/025} {\bibfield  {journal} {\bibinfo  {journal}
  {JCAP}\ }\textbf {\bibinfo {volume} {1308}},\ \bibinfo {pages} {025}
  (\bibinfo {year} {2013})},\ \Eprint {http://arxiv.org/abs/1304.4840}
  {arXiv:1304.4840 [hep-th]} \BibitemShut {NoStop}%
\bibitem [{\citenamefont {Gabadadze}\ and\ \citenamefont
  {Yu}(2016)}]{Gabadadze:2016llq}%
  \BibitemOpen
  \bibfield  {author} {\bibinfo {author} {\bibfnamefont {G.}~\bibnamefont
  {Gabadadze}}\ and\ \bibinfo {author} {\bibfnamefont {S.}~\bibnamefont {Yu}},\
  }\href@noop {} {\  (\bibinfo {year} {2016})},\ \Eprint
  {http://arxiv.org/abs/1608.01060} {arXiv:1608.01060 [hep-th]} \BibitemShut
  {NoStop}%
\bibitem [{\citenamefont {Salvatelli}\ \emph {et~al.}(2016)\citenamefont
  {Salvatelli}, \citenamefont {Piazza},\ and\ \citenamefont
  {Marinoni}}]{Salvatelli:2016mgy}%
  \BibitemOpen
  \bibfield  {author} {\bibinfo {author} {\bibfnamefont {V.}~\bibnamefont
  {Salvatelli}}, \bibinfo {author} {\bibfnamefont {F.}~\bibnamefont {Piazza}},
  \ and\ \bibinfo {author} {\bibfnamefont {C.}~\bibnamefont {Marinoni}},\
  }\href@noop {} {\  (\bibinfo {year} {2016})},\ \Eprint
  {http://arxiv.org/abs/1602.08283} {arXiv:1602.08283 [astro-ph.CO]}
  \BibitemShut {NoStop}%
\bibitem [{\citenamefont {Shahalam}\ \emph {et~al.}(2016)\citenamefont
  {Shahalam}, \citenamefont {Pacif},\ and\ \citenamefont
  {Myrzakulov}}]{Shahalam:2016kkg}%
  \BibitemOpen
  \bibfield  {author} {\bibinfo {author} {\bibfnamefont {M.}~\bibnamefont
  {Shahalam}}, \bibinfo {author} {\bibfnamefont {S.~K.~J.}\ \bibnamefont
  {Pacif}}, \ and\ \bibinfo {author} {\bibfnamefont {R.}~\bibnamefont
  {Myrzakulov}},\ }\href {\doibase 10.1140/epjc/s10052-016-4254-y} {\bibfield
  {journal} {\bibinfo  {journal} {Eur. Phys. J.}\ }\textbf {\bibinfo {volume}
  {C76}},\ \bibinfo {pages} {410} (\bibinfo {year} {2016})},\ \Eprint
  {http://arxiv.org/abs/1602.03176} {arXiv:1602.03176 [gr-qc]} \BibitemShut
  {NoStop}%
\bibitem [{\citenamefont {Minamitsuji}(2016)}]{Minamitsuji:2016qyc}%
  \BibitemOpen
  \bibfield  {author} {\bibinfo {author} {\bibfnamefont {M.}~\bibnamefont
  {Minamitsuji}},\ }\href {\doibase 10.1007/s10714-016-2025-6} {\bibfield
  {journal} {\bibinfo  {journal} {Gen. Rel. Grav.}\ }\textbf {\bibinfo {volume}
  {48}},\ \bibinfo {pages} {26} (\bibinfo {year} {2016})}\BibitemShut {NoStop}%
\bibitem [{\citenamefont {Saridakis}\ and\ \citenamefont
  {Tsoukalas}(2016{\natexlab{a}})}]{Saridakis:2016ahq}%
  \BibitemOpen
  \bibfield  {author} {\bibinfo {author} {\bibfnamefont {E.~N.}\ \bibnamefont
  {Saridakis}}\ and\ \bibinfo {author} {\bibfnamefont {M.}~\bibnamefont
  {Tsoukalas}},\ }\href {\doibase 10.1103/PhysRevD.93.124032} {\bibfield
  {journal} {\bibinfo  {journal} {Phys. Rev.}\ }\textbf {\bibinfo {volume}
  {D93}},\ \bibinfo {pages} {124032} (\bibinfo {year} {2016}{\natexlab{a}})},\
  \Eprint {http://arxiv.org/abs/1601.06734} {arXiv:1601.06734 [gr-qc]}
  \BibitemShut {NoStop}%
\bibitem [{\citenamefont {Biswas}\ and\ \citenamefont
  {Debnath}(2016)}]{Biswas:2016bwq}%
  \BibitemOpen
  \bibfield  {author} {\bibinfo {author} {\bibfnamefont {M.}~\bibnamefont
  {Biswas}}\ and\ \bibinfo {author} {\bibfnamefont {U.}~\bibnamefont
  {Debnath}},\ }\href {\doibase 10.1088/0253-6102/65/1/121} {\bibfield
  {journal} {\bibinfo  {journal} {Commun. Theor. Phys.}\ }\textbf {\bibinfo
  {volume} {65}},\ \bibinfo {pages} {121} (\bibinfo {year} {2016})}\BibitemShut
  {NoStop}%
\bibitem [{\citenamefont {Creminelli}\ \emph {et~al.}(2010)\citenamefont
  {Creminelli}, \citenamefont {Nicolis},\ and\ \citenamefont
  {Trincherini}}]{Creminelli:2010ba}%
  \BibitemOpen
  \bibfield  {author} {\bibinfo {author} {\bibfnamefont {P.}~\bibnamefont
  {Creminelli}}, \bibinfo {author} {\bibfnamefont {A.}~\bibnamefont {Nicolis}},
  \ and\ \bibinfo {author} {\bibfnamefont {E.}~\bibnamefont {Trincherini}},\
  }\href {\doibase 10.1088/1475-7516/2010/11/021} {\bibfield  {journal}
  {\bibinfo  {journal} {JCAP}\ }\textbf {\bibinfo {volume} {1011}},\ \bibinfo
  {pages} {021} (\bibinfo {year} {2010})},\ \Eprint
  {http://arxiv.org/abs/1007.0027} {arXiv:1007.0027 [hep-th]} \BibitemShut
  {NoStop}%
\bibitem [{\citenamefont {Kobayashi}\ \emph {et~al.}(2010)\citenamefont
  {Kobayashi}, \citenamefont {Yamaguchi},\ and\ \citenamefont
  {Yokoyama}}]{Kobayashi:2010cm}%
  \BibitemOpen
  \bibfield  {author} {\bibinfo {author} {\bibfnamefont {T.}~\bibnamefont
  {Kobayashi}}, \bibinfo {author} {\bibfnamefont {M.}~\bibnamefont
  {Yamaguchi}}, \ and\ \bibinfo {author} {\bibfnamefont {J.}~\bibnamefont
  {Yokoyama}},\ }\href {\doibase 10.1103/PhysRevLett.105.231302} {\bibfield
  {journal} {\bibinfo  {journal} {Phys. Rev. Lett.}\ }\textbf {\bibinfo
  {volume} {105}},\ \bibinfo {pages} {231302} (\bibinfo {year} {2010})},\
  \Eprint {http://arxiv.org/abs/1008.0603} {arXiv:1008.0603 [hep-th]}
  \BibitemShut {NoStop}%
\bibitem [{\citenamefont {Mizuno}\ and\ \citenamefont
  {Koyama}(2010)}]{Mizuno:2010ag}%
  \BibitemOpen
  \bibfield  {author} {\bibinfo {author} {\bibfnamefont {S.}~\bibnamefont
  {Mizuno}}\ and\ \bibinfo {author} {\bibfnamefont {K.}~\bibnamefont
  {Koyama}},\ }\href {\doibase 10.1103/PhysRevD.82.103518} {\bibfield
  {journal} {\bibinfo  {journal} {Phys. Rev.}\ }\textbf {\bibinfo {volume}
  {D82}},\ \bibinfo {pages} {103518} (\bibinfo {year} {2010})},\ \Eprint
  {http://arxiv.org/abs/1009.0677} {arXiv:1009.0677 [hep-th]} \BibitemShut
  {NoStop}%
\bibitem [{\citenamefont {Burrage}\ \emph {et~al.}(2011)\citenamefont
  {Burrage}, \citenamefont {de~Rham}, \citenamefont {Seery},\ and\
  \citenamefont {Tolley}}]{Burrage:2010cu}%
  \BibitemOpen
  \bibfield  {author} {\bibinfo {author} {\bibfnamefont {C.}~\bibnamefont
  {Burrage}}, \bibinfo {author} {\bibfnamefont {C.}~\bibnamefont {de~Rham}},
  \bibinfo {author} {\bibfnamefont {D.}~\bibnamefont {Seery}}, \ and\ \bibinfo
  {author} {\bibfnamefont {A.~J.}\ \bibnamefont {Tolley}},\ }\href {\doibase
  10.1088/1475-7516/2011/01/014} {\bibfield  {journal} {\bibinfo  {journal}
  {JCAP}\ }\textbf {\bibinfo {volume} {1101}},\ \bibinfo {pages} {014}
  (\bibinfo {year} {2011})},\ \Eprint {http://arxiv.org/abs/1009.2497}
  {arXiv:1009.2497 [hep-th]} \BibitemShut {NoStop}%
\bibitem [{\citenamefont {Creminelli}\ \emph {et~al.}(2011)\citenamefont
  {Creminelli}, \citenamefont {D'Amico}, \citenamefont {Musso}, \citenamefont
  {Norena},\ and\ \citenamefont {Trincherini}}]{Creminelli:2010qf}%
  \BibitemOpen
  \bibfield  {author} {\bibinfo {author} {\bibfnamefont {P.}~\bibnamefont
  {Creminelli}}, \bibinfo {author} {\bibfnamefont {G.}~\bibnamefont {D'Amico}},
  \bibinfo {author} {\bibfnamefont {M.}~\bibnamefont {Musso}}, \bibinfo
  {author} {\bibfnamefont {J.}~\bibnamefont {Norena}}, \ and\ \bibinfo {author}
  {\bibfnamefont {E.}~\bibnamefont {Trincherini}},\ }\href {\doibase
  10.1088/1475-7516/2011/02/006} {\bibfield  {journal} {\bibinfo  {journal}
  {JCAP}\ }\textbf {\bibinfo {volume} {1102}},\ \bibinfo {pages} {006}
  (\bibinfo {year} {2011})},\ \Eprint {http://arxiv.org/abs/1011.3004}
  {arXiv:1011.3004 [hep-th]} \BibitemShut {NoStop}%
\bibitem [{\citenamefont {Kamada}\ \emph {et~al.}(2011)\citenamefont {Kamada},
  \citenamefont {Kobayashi}, \citenamefont {Yamaguchi},\ and\ \citenamefont
  {Yokoyama}}]{Kamada:2010qe}%
  \BibitemOpen
  \bibfield  {author} {\bibinfo {author} {\bibfnamefont {K.}~\bibnamefont
  {Kamada}}, \bibinfo {author} {\bibfnamefont {T.}~\bibnamefont {Kobayashi}},
  \bibinfo {author} {\bibfnamefont {M.}~\bibnamefont {Yamaguchi}}, \ and\
  \bibinfo {author} {\bibfnamefont {J.}~\bibnamefont {Yokoyama}},\ }\href
  {\doibase 10.1103/PhysRevD.83.083515} {\bibfield  {journal} {\bibinfo
  {journal} {Phys. Rev.}\ }\textbf {\bibinfo {volume} {D83}},\ \bibinfo {pages}
  {083515} (\bibinfo {year} {2011})},\ \Eprint {http://arxiv.org/abs/1012.4238}
  {arXiv:1012.4238 [astro-ph.CO]} \BibitemShut {NoStop}%
\bibitem [{\citenamefont {Libanov}\ \emph {et~al.}(2016)\citenamefont
  {Libanov}, \citenamefont {Mironov},\ and\ \citenamefont
  {Rubakov}}]{Libanov:2016kfc}%
  \BibitemOpen
  \bibfield  {author} {\bibinfo {author} {\bibfnamefont {M.}~\bibnamefont
  {Libanov}}, \bibinfo {author} {\bibfnamefont {S.}~\bibnamefont {Mironov}}, \
  and\ \bibinfo {author} {\bibfnamefont {V.}~\bibnamefont {Rubakov}},\ }\href
  {\doibase 10.1088/1475-7516/2016/08/037} {\bibfield  {journal} {\bibinfo
  {journal} {JCAP}\ }\textbf {\bibinfo {volume} {1608}},\ \bibinfo {pages}
  {037} (\bibinfo {year} {2016})},\ \Eprint {http://arxiv.org/abs/1605.05992}
  {arXiv:1605.05992 [hep-th]} \BibitemShut {NoStop}%
\bibitem [{\citenamefont {Banerjee}\ and\ \citenamefont
  {Saridakis}(2016)}]{Banerjee:2016hom}%
  \BibitemOpen
  \bibfield  {author} {\bibinfo {author} {\bibfnamefont {S.}~\bibnamefont
  {Banerjee}}\ and\ \bibinfo {author} {\bibfnamefont {E.~N.}\ \bibnamefont
  {Saridakis}},\ }\href@noop {} {\  (\bibinfo {year} {2016})},\ \Eprint
  {http://arxiv.org/abs/1604.06932} {arXiv:1604.06932 [gr-qc]} \BibitemShut
  {NoStop}%
\bibitem [{\citenamefont {Hirano}\ \emph {et~al.}(2016)\citenamefont {Hirano},
  \citenamefont {Kobayashi},\ and\ \citenamefont {Yokoyama}}]{Hirano:2016gmv}%
  \BibitemOpen
  \bibfield  {author} {\bibinfo {author} {\bibfnamefont {S.}~\bibnamefont
  {Hirano}}, \bibinfo {author} {\bibfnamefont {T.}~\bibnamefont {Kobayashi}}, \
  and\ \bibinfo {author} {\bibfnamefont {S.}~\bibnamefont {Yokoyama}},\
  }\href@noop {} {\  (\bibinfo {year} {2016})},\ \Eprint
  {http://arxiv.org/abs/1604.00141} {arXiv:1604.00141 [astro-ph.CO]}
  \BibitemShut {NoStop}%
\bibitem [{\citenamefont {Brandenberger}\ and\ \citenamefont
  {Peter}(2016)}]{Brandenberger:2016vhg}%
  \BibitemOpen
  \bibfield  {author} {\bibinfo {author} {\bibfnamefont {R.}~\bibnamefont
  {Brandenberger}}\ and\ \bibinfo {author} {\bibfnamefont {P.}~\bibnamefont
  {Peter}},\ }\href@noop {} {\  (\bibinfo {year} {2016})},\ \Eprint
  {http://arxiv.org/abs/1603.05834} {arXiv:1603.05834 [hep-th]} \BibitemShut
  {NoStop}%
\bibitem [{\citenamefont {Nishi}\ and\ \citenamefont
  {Kobayashi}(2016)}]{Nishi:2016wty}%
  \BibitemOpen
  \bibfield  {author} {\bibinfo {author} {\bibfnamefont {S.}~\bibnamefont
  {Nishi}}\ and\ \bibinfo {author} {\bibfnamefont {T.}~\bibnamefont
  {Kobayashi}},\ }\href {\doibase 10.1088/1475-7516/2016/04/018} {\bibfield
  {journal} {\bibinfo  {journal} {JCAP}\ }\textbf {\bibinfo {volume} {1604}},\
  \bibinfo {pages} {018} (\bibinfo {year} {2016})},\ \Eprint
  {http://arxiv.org/abs/1601.06561} {arXiv:1601.06561 [hep-th]} \BibitemShut
  {NoStop}%
\bibitem [{\citenamefont {Neveu}\ \emph {et~al.}(2016)\citenamefont {Neveu},
  \citenamefont {Ruhlmann-Kleider}, \citenamefont {Astier}, \citenamefont
  {Besancon}, \citenamefont {Guy}, \citenamefont {M$\ddot{o}$ller},\ and\
  \citenamefont {Babichev}}]{Neveu:2016gxp}%
  \BibitemOpen
  \bibfield  {author} {\bibinfo {author} {\bibfnamefont {J.}~\bibnamefont
  {Neveu}}, \bibinfo {author} {\bibfnamefont {V.}~\bibnamefont
  {Ruhlmann-Kleider}}, \bibinfo {author} {\bibfnamefont {P.}~\bibnamefont
  {Astier}}, \bibinfo {author} {\bibfnamefont {M.}~\bibnamefont {Besancon}},
  \bibinfo {author} {\bibfnamefont {J.}~\bibnamefont {Guy}}, \bibinfo {author}
  {\bibfnamefont {A.}~\bibnamefont {M$\ddot{o}$ller}}, \ and\ \bibinfo {author}
  {\bibfnamefont {E.}~\bibnamefont {Babichev}},\ }\href@noop {} {\bibfield
  {journal} {\bibinfo  {journal} {arXiv:1605.02627 [gr-qc]}\ } (\bibinfo {year}
  {2016})},\ \Eprint {http://arxiv.org/abs/1605.02627} {arXiv:1605.02627
  [gr-qc]} \BibitemShut {NoStop}%
\bibitem [{\citenamefont {Deffayet}\ \emph
  {et~al.}(2010{\natexlab{b}})\citenamefont {Deffayet}, \citenamefont {Deser},\
  and\ \citenamefont {Esposito-Farese}}]{Deffayet:2010zh}%
  \BibitemOpen
  \bibfield  {author} {\bibinfo {author} {\bibfnamefont {C.}~\bibnamefont
  {Deffayet}}, \bibinfo {author} {\bibfnamefont {S.}~\bibnamefont {Deser}}, \
  and\ \bibinfo {author} {\bibfnamefont {G.}~\bibnamefont {Esposito-Farese}},\
  }\href {\doibase 10.1103/PhysRevD.82.061501} {\bibfield  {journal} {\bibinfo
  {journal} {Phys. Rev.}\ }\textbf {\bibinfo {volume} {D82}},\ \bibinfo {pages}
  {061501} (\bibinfo {year} {2010}{\natexlab{b}})},\ \Eprint
  {http://arxiv.org/abs/1007.5278} {arXiv:1007.5278 [gr-qc]} \BibitemShut
  {NoStop}%
\bibitem [{\citenamefont {Heisenberg}(2014)}]{Heisenberg:2014rta}%
  \BibitemOpen
  \bibfield  {author} {\bibinfo {author} {\bibfnamefont {L.}~\bibnamefont
  {Heisenberg}},\ }\href {\doibase 10.1088/1475-7516/2014/05/015} {\bibfield
  {journal} {\bibinfo  {journal} {JCAP}\ }\textbf {\bibinfo {volume} {1405}},\
  \bibinfo {pages} {015} (\bibinfo {year} {2014})},\ \Eprint
  {http://arxiv.org/abs/1402.7026} {arXiv:1402.7026 [hep-th]} \BibitemShut
  {NoStop}%
\bibitem [{\citenamefont {Tasinato}(2014{\natexlab{a}})}]{Tasinato:2014eka}%
  \BibitemOpen
  \bibfield  {author} {\bibinfo {author} {\bibfnamefont {G.}~\bibnamefont
  {Tasinato}},\ }\href {\doibase 10.1007/JHEP04(2014)067} {\bibfield  {journal}
  {\bibinfo  {journal} {JHEP}\ }\textbf {\bibinfo {volume} {1404}},\ \bibinfo
  {pages} {067} (\bibinfo {year} {2014}{\natexlab{a}})},\ \Eprint
  {http://arxiv.org/abs/1402.6450} {arXiv:1402.6450 [hep-th]} \BibitemShut
  {NoStop}%
\bibitem [{\citenamefont {Allys}\ \emph
  {et~al.}(2016{\natexlab{a}})\citenamefont {Allys}, \citenamefont {Peter},\
  and\ \citenamefont {Rodr\'{\i}guez}}]{Allys:2015sht}%
  \BibitemOpen
  \bibfield  {author} {\bibinfo {author} {\bibfnamefont {E.}~\bibnamefont
  {Allys}}, \bibinfo {author} {\bibfnamefont {P.}~\bibnamefont {Peter}}, \ and\
  \bibinfo {author} {\bibfnamefont {Y.}~\bibnamefont {Rodr\'{\i}guez}},\ }\href
  {\doibase 10.1088/1475-7516/2016/02/004} {\bibfield  {journal} {\bibinfo
  {journal} {JCAP}\ }\textbf {\bibinfo {volume} {1602}},\ \bibinfo {pages}
  {004} (\bibinfo {year} {2016}{\natexlab{a}})},\ \Eprint
  {http://arxiv.org/abs/1511.03101} {arXiv:1511.03101 [hep-th]} \BibitemShut
  {NoStop}%
\bibitem [{\citenamefont {Beltr\'an~Jimenez}\ and\ \citenamefont
  {Heisenberg}(2016)}]{Jimenez:2016isa}%
  \BibitemOpen
  \bibfield  {author} {\bibinfo {author} {\bibfnamefont {J.}~\bibnamefont
  {Beltr\'an~Jimenez}}\ and\ \bibinfo {author} {\bibfnamefont {L.}~\bibnamefont
  {Heisenberg}},\ }\href {\doibase 10.1016/j.physletb.2016.04.017} {\bibfield
  {journal} {\bibinfo  {journal} {Phys. Lett.}\ }\textbf {\bibinfo {volume}
  {B757}},\ \bibinfo {pages} {405} (\bibinfo {year} {2016})},\ \Eprint
  {http://arxiv.org/abs/1602.03410} {arXiv:1602.03410 [hep-th]} \BibitemShut
  {NoStop}%
\bibitem [{\citenamefont {Allys}\ \emph
  {et~al.}(2016{\natexlab{b}})\citenamefont {Allys}, \citenamefont
  {Beltran~Almeida}, \citenamefont {Peter},\ and\ \citenamefont
  {Rodríguez}}]{Allys:2016jaq}%
  \BibitemOpen
  \bibfield  {author} {\bibinfo {author} {\bibfnamefont {E.}~\bibnamefont
  {Allys}}, \bibinfo {author} {\bibfnamefont {J.~P.}\ \bibnamefont
  {Beltran~Almeida}}, \bibinfo {author} {\bibfnamefont {P.}~\bibnamefont
  {Peter}}, \ and\ \bibinfo {author} {\bibfnamefont {Y.}~\bibnamefont
  {Rodríguez}},\ }\href {\doibase 10.1088/1475-7516/2016/09/026} {\bibfield
  {journal} {\bibinfo  {journal} {JCAP}\ }\textbf {\bibinfo {volume} {1609}},\
  \bibinfo {pages} {026} (\bibinfo {year} {2016}{\natexlab{b}})},\ \Eprint
  {http://arxiv.org/abs/1605.08355} {arXiv:1605.08355 [hep-th]} \BibitemShut
  {NoStop}%
\bibitem [{\citenamefont {Tasinato}(2014{\natexlab{b}})}]{Tasinato:2014mia}%
  \BibitemOpen
  \bibfield  {author} {\bibinfo {author} {\bibfnamefont {G.}~\bibnamefont
  {Tasinato}},\ }\href {\doibase 10.1088/0264-9381/31/22/225004} {\bibfield
  {journal} {\bibinfo  {journal} {Class. Quant. Grav.}\ }\textbf {\bibinfo
  {volume} {31}},\ \bibinfo {pages} {225004} (\bibinfo {year}
  {2014}{\natexlab{b}})},\ \Eprint {http://arxiv.org/abs/1404.4883}
  {arXiv:1404.4883 [hep-th]} \BibitemShut {NoStop}%
\bibitem [{\citenamefont {Hull}\ \emph {et~al.}(2015)\citenamefont {Hull},
  \citenamefont {Koyama},\ and\ \citenamefont {Tasinato}}]{Hull:2014bga}%
  \BibitemOpen
  \bibfield  {author} {\bibinfo {author} {\bibfnamefont {M.}~\bibnamefont
  {Hull}}, \bibinfo {author} {\bibfnamefont {K.}~\bibnamefont {Koyama}}, \ and\
  \bibinfo {author} {\bibfnamefont {G.}~\bibnamefont {Tasinato}},\ }\href
  {\doibase 10.1007/JHEP03(2015)154} {\bibfield  {journal} {\bibinfo  {journal}
  {JHEP}\ }\textbf {\bibinfo {volume} {1503}},\ \bibinfo {pages} {154}
  (\bibinfo {year} {2015})},\ \Eprint {http://arxiv.org/abs/1408.6871}
  {arXiv:1408.6871 [hep-th]} \BibitemShut {NoStop}%
\bibitem [{\citenamefont {De~Felice}\ \emph
  {et~al.}(2016{\natexlab{a}})\citenamefont {De~Felice}, \citenamefont
  {Heisenberg}, \citenamefont {Kase}, \citenamefont {Tsujikawa}, \citenamefont
  {Zhang},\ and\ \citenamefont {Zhao}}]{DeFelice:2016cri}%
  \BibitemOpen
  \bibfield  {author} {\bibinfo {author} {\bibfnamefont {A.}~\bibnamefont
  {De~Felice}}, \bibinfo {author} {\bibfnamefont {L.}~\bibnamefont
  {Heisenberg}}, \bibinfo {author} {\bibfnamefont {R.}~\bibnamefont {Kase}},
  \bibinfo {author} {\bibfnamefont {S.}~\bibnamefont {Tsujikawa}}, \bibinfo
  {author} {\bibfnamefont {Y.-l.}\ \bibnamefont {Zhang}}, \ and\ \bibinfo
  {author} {\bibfnamefont {G.-B.}\ \bibnamefont {Zhao}},\ }\href {\doibase
  10.1103/PhysRevD.93.104016} {\bibfield  {journal} {\bibinfo  {journal} {Phys.
  Rev.}\ }\textbf {\bibinfo {volume} {D93}},\ \bibinfo {pages} {104016}
  (\bibinfo {year} {2016}{\natexlab{a}})},\ \Eprint
  {http://arxiv.org/abs/1602.00371} {arXiv:1602.00371 [gr-qc]} \BibitemShut
  {NoStop}%
\bibitem [{\citenamefont {De~Felice}\ \emph
  {et~al.}(2016{\natexlab{b}})\citenamefont {De~Felice}, \citenamefont
  {Heisenberg}, \citenamefont {Kase}, \citenamefont {Mukohyama}, \citenamefont
  {Tsujikawa},\ and\ \citenamefont {Zhang}}]{DeFelice:2016yws}%
  \BibitemOpen
  \bibfield  {author} {\bibinfo {author} {\bibfnamefont {A.}~\bibnamefont
  {De~Felice}}, \bibinfo {author} {\bibfnamefont {L.}~\bibnamefont
  {Heisenberg}}, \bibinfo {author} {\bibfnamefont {R.}~\bibnamefont {Kase}},
  \bibinfo {author} {\bibfnamefont {S.}~\bibnamefont {Mukohyama}}, \bibinfo
  {author} {\bibfnamefont {S.}~\bibnamefont {Tsujikawa}}, \ and\ \bibinfo
  {author} {\bibfnamefont {Y.-l.}\ \bibnamefont {Zhang}},\ }\href {\doibase
  10.1088/1475-7516/2016/06/048} {\bibfield  {journal} {\bibinfo  {journal}
  {JCAP}\ }\textbf {\bibinfo {volume} {1606}},\ \bibinfo {pages} {048}
  (\bibinfo {year} {2016}{\natexlab{b}})},\ \Eprint
  {http://arxiv.org/abs/1603.05806} {arXiv:1603.05806 [gr-qc]} \BibitemShut
  {NoStop}%
\bibitem [{\citenamefont {De~Felice}\ \emph
  {et~al.}(2016{\natexlab{c}})\citenamefont {De~Felice}, \citenamefont
  {Heisenberg}, \citenamefont {Kase}, \citenamefont {Mukohyama}, \citenamefont
  {Tsujikawa},\ and\ \citenamefont {Zhang}}]{DeFelice:2016uil}%
  \BibitemOpen
  \bibfield  {author} {\bibinfo {author} {\bibfnamefont {A.}~\bibnamefont
  {De~Felice}}, \bibinfo {author} {\bibfnamefont {L.}~\bibnamefont
  {Heisenberg}}, \bibinfo {author} {\bibfnamefont {R.}~\bibnamefont {Kase}},
  \bibinfo {author} {\bibfnamefont {S.}~\bibnamefont {Mukohyama}}, \bibinfo
  {author} {\bibfnamefont {S.}~\bibnamefont {Tsujikawa}}, \ and\ \bibinfo
  {author} {\bibfnamefont {Y.-l.}\ \bibnamefont {Zhang}},\ }\href {\doibase
  10.1103/PhysRevD.94.044024} {\bibfield  {journal} {\bibinfo  {journal} {Phys.
  Rev.}\ }\textbf {\bibinfo {volume} {D94}},\ \bibinfo {pages} {044024}
  (\bibinfo {year} {2016}{\natexlab{c}})},\ \Eprint
  {http://arxiv.org/abs/1605.05066} {arXiv:1605.05066 [gr-qc]} \BibitemShut
  {NoStop}%
\bibitem [{\citenamefont {Heisenberg}\ \emph {et~al.}(2016)\citenamefont
  {Heisenberg}, \citenamefont {Kase},\ and\ \citenamefont
  {Tsujikawa}}]{Heisenberg:2016wtr}%
  \BibitemOpen
  \bibfield  {author} {\bibinfo {author} {\bibfnamefont {L.}~\bibnamefont
  {Heisenberg}}, \bibinfo {author} {\bibfnamefont {R.}~\bibnamefont {Kase}}, \
  and\ \bibinfo {author} {\bibfnamefont {S.}~\bibnamefont {Tsujikawa}},\
  }\href@noop {} {\  (\bibinfo {year} {2016})},\ \Eprint
  {http://arxiv.org/abs/1607.03175} {arXiv:1607.03175 [gr-qc]} \BibitemShut
  {NoStop}%
\bibitem [{\citenamefont {Padilla}\ \emph {et~al.}(2010)\citenamefont
  {Padilla}, \citenamefont {Saffin},\ and\ \citenamefont
  {Zhou}}]{Padilla:2010de}%
  \BibitemOpen
  \bibfield  {author} {\bibinfo {author} {\bibfnamefont {A.}~\bibnamefont
  {Padilla}}, \bibinfo {author} {\bibfnamefont {P.~M.}\ \bibnamefont {Saffin}},
  \ and\ \bibinfo {author} {\bibfnamefont {S.-Y.}\ \bibnamefont {Zhou}},\
  }\href {\doibase 10.1007/JHEP12(2010)031} {\bibfield  {journal} {\bibinfo
  {journal} {JHEP}\ }\textbf {\bibinfo {volume} {1012}},\ \bibinfo {pages}
  {031} (\bibinfo {year} {2010})},\ \Eprint {http://arxiv.org/abs/1007.5424}
  {arXiv:1007.5424 [hep-th]} \BibitemShut {NoStop}%
\bibitem [{\citenamefont {Padilla}\ \emph
  {et~al.}(2011{\natexlab{a}})\citenamefont {Padilla}, \citenamefont {Saffin},\
  and\ \citenamefont {Zhou}}]{Padilla:2010tj}%
  \BibitemOpen
  \bibfield  {author} {\bibinfo {author} {\bibfnamefont {A.}~\bibnamefont
  {Padilla}}, \bibinfo {author} {\bibfnamefont {P.~M.}\ \bibnamefont {Saffin}},
  \ and\ \bibinfo {author} {\bibfnamefont {S.-Y.}\ \bibnamefont {Zhou}},\
  }\href {\doibase 10.1007/JHEP01(2011)099} {\bibfield  {journal} {\bibinfo
  {journal} {JHEP}\ }\textbf {\bibinfo {volume} {1101}},\ \bibinfo {pages}
  {099} (\bibinfo {year} {2011}{\natexlab{a}})},\ \Eprint
  {http://arxiv.org/abs/1008.3312} {arXiv:1008.3312 [hep-th]} \BibitemShut
  {NoStop}%
\bibitem [{\citenamefont {Padilla}\ \emph
  {et~al.}(2011{\natexlab{b}})\citenamefont {Padilla}, \citenamefont {Saffin},\
  and\ \citenamefont {Zhou}}]{Padilla:2010ir}%
  \BibitemOpen
  \bibfield  {author} {\bibinfo {author} {\bibfnamefont {A.}~\bibnamefont
  {Padilla}}, \bibinfo {author} {\bibfnamefont {P.~M.}\ \bibnamefont {Saffin}},
  \ and\ \bibinfo {author} {\bibfnamefont {S.-Y.}\ \bibnamefont {Zhou}},\
  }\href {\doibase 10.1103/PhysRevD.83.045009} {\bibfield  {journal} {\bibinfo
  {journal} {Phys. Rev.}\ }\textbf {\bibinfo {volume} {D83}},\ \bibinfo {pages}
  {045009} (\bibinfo {year} {2011}{\natexlab{b}})},\ \Eprint
  {http://arxiv.org/abs/1008.0745} {arXiv:1008.0745 [hep-th]} \BibitemShut
  {NoStop}%
\bibitem [{\citenamefont {Zhou}(2011)}]{Zhou:2010di}%
  \BibitemOpen
  \bibfield  {author} {\bibinfo {author} {\bibfnamefont {S.-Y.}\ \bibnamefont
  {Zhou}},\ }\href {\doibase 10.1103/PhysRevD.83.064005} {\bibfield  {journal}
  {\bibinfo  {journal} {Phys. Rev.}\ }\textbf {\bibinfo {volume} {D83}},\
  \bibinfo {pages} {064005} (\bibinfo {year} {2011})},\ \Eprint
  {http://arxiv.org/abs/1011.0863} {arXiv:1011.0863 [hep-th]} \BibitemShut
  {NoStop}%
\bibitem [{\citenamefont {Hinterbichler}\ \emph {et~al.}(2010)\citenamefont
  {Hinterbichler}, \citenamefont {Trodden},\ and\ \citenamefont
  {Wesley}}]{Hinterbichler:2010xn}%
  \BibitemOpen
  \bibfield  {author} {\bibinfo {author} {\bibfnamefont {K.}~\bibnamefont
  {Hinterbichler}}, \bibinfo {author} {\bibfnamefont {M.}~\bibnamefont
  {Trodden}}, \ and\ \bibinfo {author} {\bibfnamefont {D.}~\bibnamefont
  {Wesley}},\ }\href {\doibase 10.1103/PhysRevD.82.124018} {\bibfield
  {journal} {\bibinfo  {journal} {Phys. Rev.}\ }\textbf {\bibinfo {volume}
  {D82}},\ \bibinfo {pages} {124018} (\bibinfo {year} {2010})},\ \Eprint
  {http://arxiv.org/abs/1008.1305} {arXiv:1008.1305 [hep-th]} \BibitemShut
  {NoStop}%
\bibitem [{\citenamefont {Andrews}\ \emph {et~al.}(2011)\citenamefont
  {Andrews}, \citenamefont {Hinterbichler}, \citenamefont {Khoury},\ and\
  \citenamefont {Trodden}}]{Andrews:2010km}%
  \BibitemOpen
  \bibfield  {author} {\bibinfo {author} {\bibfnamefont {M.}~\bibnamefont
  {Andrews}}, \bibinfo {author} {\bibfnamefont {K.}~\bibnamefont
  {Hinterbichler}}, \bibinfo {author} {\bibfnamefont {J.}~\bibnamefont
  {Khoury}}, \ and\ \bibinfo {author} {\bibfnamefont {M.}~\bibnamefont
  {Trodden}},\ }\href {\doibase 10.1103/PhysRevD.83.044042} {\bibfield
  {journal} {\bibinfo  {journal} {Phys. Rev.}\ }\textbf {\bibinfo {volume}
  {D83}},\ \bibinfo {pages} {044042} (\bibinfo {year} {2011})},\ \Eprint
  {http://arxiv.org/abs/1008.4128} {arXiv:1008.4128 [hep-th]} \BibitemShut
  {NoStop}%
\bibitem [{\citenamefont {Padilla}\ and\ \citenamefont
  {Sivanesan}(2013)}]{Padilla:2012dx}%
  \BibitemOpen
  \bibfield  {author} {\bibinfo {author} {\bibfnamefont {A.}~\bibnamefont
  {Padilla}}\ and\ \bibinfo {author} {\bibfnamefont {V.}~\bibnamefont
  {Sivanesan}},\ }\href {\doibase 10.1007/JHEP04(2013)032} {\bibfield
  {journal} {\bibinfo  {journal} {JHEP}\ }\textbf {\bibinfo {volume} {1304}},\
  \bibinfo {pages} {032} (\bibinfo {year} {2013})},\ \Eprint
  {http://arxiv.org/abs/1210.4026} {arXiv:1210.4026 [gr-qc]} \BibitemShut
  {NoStop}%
\bibitem [{\citenamefont {Sivanesan}(2014)}]{Sivanesan:2013tba}%
  \BibitemOpen
  \bibfield  {author} {\bibinfo {author} {\bibfnamefont {V.}~\bibnamefont
  {Sivanesan}},\ }\href {\doibase 10.1103/PhysRevD.90.104006} {\bibfield
  {journal} {\bibinfo  {journal} {Phys. Rev.}\ }\textbf {\bibinfo {volume}
  {D90}},\ \bibinfo {pages} {104006} (\bibinfo {year} {2014})},\ \Eprint
  {http://arxiv.org/abs/1307.8081} {arXiv:1307.8081 [gr-qc]} \BibitemShut
  {NoStop}%
\bibitem [{\citenamefont {Garcia-Saenz}(2013)}]{Garcia-Saenz:2013gya}%
  \BibitemOpen
  \bibfield  {author} {\bibinfo {author} {\bibfnamefont {S.}~\bibnamefont
  {Garcia-Saenz}},\ }\href {\doibase 10.1103/PhysRevD.87.104012} {\bibfield
  {journal} {\bibinfo  {journal} {Phys. Rev.}\ }\textbf {\bibinfo {volume}
  {D87}},\ \bibinfo {pages} {104012} (\bibinfo {year} {2013})},\ \Eprint
  {http://arxiv.org/abs/1303.2905} {arXiv:1303.2905 [hep-th]} \BibitemShut
  {NoStop}%
\bibitem [{\citenamefont {Charmousis}\ \emph {et~al.}(2014)\citenamefont
  {Charmousis}, \citenamefont {Kolyvaris}, \citenamefont {Papantonopoulos},\
  and\ \citenamefont {Tsoukalas}}]{Charmousis:2014zaa}%
  \BibitemOpen
  \bibfield  {author} {\bibinfo {author} {\bibfnamefont {C.}~\bibnamefont
  {Charmousis}}, \bibinfo {author} {\bibfnamefont {T.}~\bibnamefont
  {Kolyvaris}}, \bibinfo {author} {\bibfnamefont {E.}~\bibnamefont
  {Papantonopoulos}}, \ and\ \bibinfo {author} {\bibfnamefont {M.}~\bibnamefont
  {Tsoukalas}},\ }\href {\doibase 10.1007/JHEP07(2014)085} {\bibfield
  {journal} {\bibinfo  {journal} {JHEP}\ }\textbf {\bibinfo {volume} {07}},\
  \bibinfo {pages} {085} (\bibinfo {year} {2014})},\ \Eprint
  {http://arxiv.org/abs/1404.1024} {arXiv:1404.1024 [gr-qc]} \BibitemShut
  {NoStop}%
\bibitem [{\citenamefont {Saridakis}\ and\ \citenamefont
  {Tsoukalas}(2016{\natexlab{b}})}]{Saridakis:2016mjd}%
  \BibitemOpen
  \bibfield  {author} {\bibinfo {author} {\bibfnamefont {E.~N.}\ \bibnamefont
  {Saridakis}}\ and\ \bibinfo {author} {\bibfnamefont {M.}~\bibnamefont
  {Tsoukalas}},\ }\href {\doibase 10.1088/1475-7516/2016/04/017} {\bibfield
  {journal} {\bibinfo  {journal} {JCAP}\ }\textbf {\bibinfo {volume} {1604}},\
  \bibinfo {pages} {017} (\bibinfo {year} {2016}{\natexlab{b}})},\ \Eprint
  {http://arxiv.org/abs/1602.06890} {arXiv:1602.06890 [gr-qc]} \BibitemShut
  {NoStop}%
\bibitem [{\citenamefont {Kobayashi}\ \emph {et~al.}(2013)\citenamefont
  {Kobayashi}, \citenamefont {Tanahashi},\ and\ \citenamefont
  {Yamaguchi}}]{Kobayashi:2013ina}%
  \BibitemOpen
  \bibfield  {author} {\bibinfo {author} {\bibfnamefont {T.}~\bibnamefont
  {Kobayashi}}, \bibinfo {author} {\bibfnamefont {N.}~\bibnamefont
  {Tanahashi}}, \ and\ \bibinfo {author} {\bibfnamefont {M.}~\bibnamefont
  {Yamaguchi}},\ }\href {\doibase 10.1103/PhysRevD.88.083504} {\bibfield
  {journal} {\bibinfo  {journal} {Phys. Rev.}\ }\textbf {\bibinfo {volume}
  {D88}},\ \bibinfo {pages} {083504} (\bibinfo {year} {2013})},\ \Eprint
  {http://arxiv.org/abs/1308.4798} {arXiv:1308.4798 [hep-th]} \BibitemShut
  {NoStop}%
\bibitem [{\citenamefont {Ohashi}\ \emph {et~al.}(2015)\citenamefont {Ohashi},
  \citenamefont {Tanahashi}, \citenamefont {Kobayashi},\ and\ \citenamefont
  {Yamaguchi}}]{Ohashi:2015fma}%
  \BibitemOpen
  \bibfield  {author} {\bibinfo {author} {\bibfnamefont {S.}~\bibnamefont
  {Ohashi}}, \bibinfo {author} {\bibfnamefont {N.}~\bibnamefont {Tanahashi}},
  \bibinfo {author} {\bibfnamefont {T.}~\bibnamefont {Kobayashi}}, \ and\
  \bibinfo {author} {\bibfnamefont {M.}~\bibnamefont {Yamaguchi}},\ }\href
  {\doibase 10.1007/JHEP07(2015)008} {\bibfield  {journal} {\bibinfo  {journal}
  {JHEP}\ }\textbf {\bibinfo {volume} {07}},\ \bibinfo {pages} {008} (\bibinfo
  {year} {2015})},\ \Eprint {http://arxiv.org/abs/1505.06029} {arXiv:1505.06029
  [gr-qc]} \BibitemShut {NoStop}%
\bibitem [{\citenamefont {Woodard}(2007)}]{Woodard:2006nt}%
  \BibitemOpen
  \bibfield  {author} {\bibinfo {author} {\bibfnamefont {R.~P.}\ \bibnamefont
  {Woodard}},\ }\bibfield  {booktitle} {\emph {\bibinfo {booktitle} {{The
  invisible universe: Dark matter and dark energy. Proceedings, 3rd Aegean
  School, Karfas, Greece, September 26-October 1, 2005}}},\ }\href {\doibase
  10.1007/978-3-540-71013-4_14} {\bibfield  {journal} {\bibinfo  {journal}
  {Lect. Notes Phys.}\ }\textbf {\bibinfo {volume} {720}},\ \bibinfo {pages}
  {403} (\bibinfo {year} {2007})},\ \Eprint
  {http://arxiv.org/abs/astro-ph/0601672} {arXiv:astro-ph/0601672 [astro-ph]}
  \BibitemShut {NoStop}%
\bibitem [{\citenamefont {Woodard}(2015)}]{Woodard:2015zca}%
  \BibitemOpen
  \bibfield  {author} {\bibinfo {author} {\bibfnamefont {R.~P.}\ \bibnamefont
  {Woodard}},\ }\href {\doibase 10.4249/scholarpedia.32243} {\bibfield
  {journal} {\bibinfo  {journal} {Scholarpedia}\ }\textbf {\bibinfo {volume}
  {10}},\ \bibinfo {pages} {32243} (\bibinfo {year} {2015})},\ \Eprint
  {http://arxiv.org/abs/1506.02210} {arXiv:1506.02210 [hep-th]} \BibitemShut
  {NoStop}%
\bibitem [{\citenamefont {Motohashi}\ and\ \citenamefont
  {Suyama}(2015)}]{Motohashi:2014opa}%
  \BibitemOpen
  \bibfield  {author} {\bibinfo {author} {\bibfnamefont {H.}~\bibnamefont
  {Motohashi}}\ and\ \bibinfo {author} {\bibfnamefont {T.}~\bibnamefont
  {Suyama}},\ }\href {\doibase 10.1103/PhysRevD.91.085009} {\bibfield
  {journal} {\bibinfo  {journal} {Phys. Rev.}\ }\textbf {\bibinfo {volume}
  {D91}},\ \bibinfo {pages} {085009} (\bibinfo {year} {2015})},\ \Eprint
  {http://arxiv.org/abs/1411.3721} {arXiv:1411.3721 [physics.class-ph]}
  \BibitemShut {NoStop}%
\bibitem [{\citenamefont {Fuchs}\ and\ \citenamefont
  {Schweigert}(2003)}]{Fuchs:1997jv}%
  \BibitemOpen
  \bibfield  {author} {\bibinfo {author} {\bibfnamefont {J.}~\bibnamefont
  {Fuchs}}\ and\ \bibinfo {author} {\bibfnamefont {C.}~\bibnamefont
  {Schweigert}},\ }\href
  {http://www.cambridge.org/uk/catalogue/catalogue.asp?isbn=0521582733} {\emph
  {\bibinfo {title} {{Symmetries, Lie algebras and representations: A graduate
  course for physicists}}}}\ (\bibinfo  {publisher} {Cambridge University
  Press},\ \bibinfo {year} {2003})\BibitemShut {NoStop}%
\bibitem [{\citenamefont {Slansky}(1981)}]{Slansky:1981yr}%
  \BibitemOpen
  \bibfield  {author} {\bibinfo {author} {\bibfnamefont {R.}~\bibnamefont
  {Slansky}},\ }\href {\doibase 10.1016/0370-1573(81)90092-2} {\bibfield
  {journal} {\bibinfo  {journal} {Phys.Rept.}\ }\textbf {\bibinfo {volume}
  {79}},\ \bibinfo {pages} {1} (\bibinfo {year} {1981})}\BibitemShut {NoStop}%
\bibitem [{\citenamefont {Ramond}(2010)}]{Ramond:2010zz}%
  \BibitemOpen
  \bibfield  {author} {\bibinfo {author} {\bibfnamefont {P.}~\bibnamefont
  {Ramond}},\ }\href {http://www.cambridge.org/de/knowledge/isbn/item2710157}
  {\emph {\bibinfo {title} {{Group theory: A physicist's survey}}}}\ (\bibinfo
  {publisher} {Cambridge University Press},\ \bibinfo {year}
  {2010})\BibitemShut {NoStop}%
\bibitem [{\citenamefont {de~Azcarraga}\ \emph {et~al.}(1998)\citenamefont
  {de~Azcarraga}, \citenamefont {Macfarlane}, \citenamefont {Mountain},\ and\
  \citenamefont {Perez~Bueno}}]{deAzcarraga:1997ya}%
  \BibitemOpen
  \bibfield  {author} {\bibinfo {author} {\bibfnamefont {J.~A.}\ \bibnamefont
  {de~Azcarraga}}, \bibinfo {author} {\bibfnamefont {A.~J.}\ \bibnamefont
  {Macfarlane}}, \bibinfo {author} {\bibfnamefont {A.~J.}\ \bibnamefont
  {Mountain}}, \ and\ \bibinfo {author} {\bibfnamefont {J.~C.}\ \bibnamefont
  {Perez~Bueno}},\ }\href {\doibase 10.1016/S0550-3213(97)00609-3} {\bibfield
  {journal} {\bibinfo  {journal} {Nucl. Phys.}\ }\textbf {\bibinfo {volume}
  {B510}},\ \bibinfo {pages} {657} (\bibinfo {year} {1998})},\ \Eprint
  {http://arxiv.org/abs/physics/9706006} {arXiv:physics/9706006 [physics]}
  \BibitemShut {NoStop}%
\bibitem [{\citenamefont {Dittner}(1972)}]{Dittner:1972hm}%
  \BibitemOpen
  \bibfield  {author} {\bibinfo {author} {\bibfnamefont {P.}~\bibnamefont
  {Dittner}},\ }\href {\doibase 10.1007/BF01649658} {\bibfield  {journal}
  {\bibinfo  {journal} {Commun. Math. Phys.}\ }\textbf {\bibinfo {volume}
  {27}},\ \bibinfo {pages} {44} (\bibinfo {year} {1972})}\BibitemShut {NoStop}%
\bibitem [{\citenamefont {Metha}\ \emph {et~al.}(1983)\citenamefont {Metha},
  \citenamefont {Normand},\ and\ \citenamefont {Gupta}}]{Metha:1983mng}%
  \BibitemOpen
  \bibfield  {author} {\bibinfo {author} {\bibfnamefont {M.~L.}\ \bibnamefont
  {Metha}}, \bibinfo {author} {\bibfnamefont {J.~M.}\ \bibnamefont {Normand}},
  \ and\ \bibinfo {author} {\bibfnamefont {V.}~\bibnamefont {Gupta}},\
  }\href@noop {} {\bibfield  {journal} {\bibinfo  {journal} {Commun. Math.
  Phys.}\ }\textbf {\bibinfo {volume} {90}},\ \bibinfo {pages} {69} (\bibinfo
  {year} {1983})}\BibitemShut {NoStop}%
\bibitem [{\citenamefont {Feger}\ and\ \citenamefont
  {Kephart}(2015)}]{Feger:2012bs}%
  \BibitemOpen
  \bibfield  {author} {\bibinfo {author} {\bibfnamefont {R.}~\bibnamefont
  {Feger}}\ and\ \bibinfo {author} {\bibfnamefont {T.~W.}\ \bibnamefont
  {Kephart}},\ }\href {\doibase 10.1016/j.cpc.2014.12.023} {\bibfield
  {journal} {\bibinfo  {journal} {Comput. Phys. Commun.}\ }\textbf {\bibinfo
  {volume} {192}},\ \bibinfo {pages} {166} (\bibinfo {year} {2015})},\ \Eprint
  {http://arxiv.org/abs/1206.6379} {arXiv:1206.6379 [math-ph]} \BibitemShut
  {NoStop}%
\bibitem [{\citenamefont {Gleyzes}\ \emph {et~al.}(2015)\citenamefont
  {Gleyzes}, \citenamefont {Langlois}, \citenamefont {Piazza},\ and\
  \citenamefont {Vernizzi}}]{Gleyzes:2014dya}%
  \BibitemOpen
  \bibfield  {author} {\bibinfo {author} {\bibfnamefont {J.}~\bibnamefont
  {Gleyzes}}, \bibinfo {author} {\bibfnamefont {D.}~\bibnamefont {Langlois}},
  \bibinfo {author} {\bibfnamefont {F.}~\bibnamefont {Piazza}}, \ and\ \bibinfo
  {author} {\bibfnamefont {F.}~\bibnamefont {Vernizzi}},\ }\href {\doibase
  10.1103/PhysRevLett.114.211101} {\bibfield  {journal} {\bibinfo  {journal}
  {Phys.Rev.Lett.}\ }\textbf {\bibinfo {volume} {114}},\ \bibinfo {pages}
  {211101} (\bibinfo {year} {2015})},\ \Eprint {http://arxiv.org/abs/1404.6495}
  {arXiv:1404.6495 [hep-th]} \BibitemShut {NoStop}%
\bibitem [{\citenamefont {Langlois}\ and\ \citenamefont
  {Noui}(2016{\natexlab{a}})}]{Langlois:2015cwa}%
  \BibitemOpen
  \bibfield  {author} {\bibinfo {author} {\bibfnamefont {D.}~\bibnamefont
  {Langlois}}\ and\ \bibinfo {author} {\bibfnamefont {K.}~\bibnamefont
  {Noui}},\ }\href {\doibase 10.1088/1475-7516/2016/02/034} {\bibfield
  {journal} {\bibinfo  {journal} {JCAP}\ }\textbf {\bibinfo {volume} {1602}},\
  \bibinfo {pages} {034} (\bibinfo {year} {2016}{\natexlab{a}})},\ \Eprint
  {http://arxiv.org/abs/1510.06930} {arXiv:1510.06930 [gr-qc]} \BibitemShut
  {NoStop}%
\bibitem [{\citenamefont {Langlois}\ and\ \citenamefont
  {Noui}(2016{\natexlab{b}})}]{Langlois:2015skt}%
  \BibitemOpen
  \bibfield  {author} {\bibinfo {author} {\bibfnamefont {D.}~\bibnamefont
  {Langlois}}\ and\ \bibinfo {author} {\bibfnamefont {K.}~\bibnamefont
  {Noui}},\ }\href {\doibase 10.1088/1475-7516/2016/07/016} {\bibfield
  {journal} {\bibinfo  {journal} {JCAP}\ }\textbf {\bibinfo {volume} {1607}},\
  \bibinfo {pages} {016} (\bibinfo {year} {2016}{\natexlab{b}})},\ \Eprint
  {http://arxiv.org/abs/1512.06820} {arXiv:1512.06820 [gr-qc]} \BibitemShut
  {NoStop}%
\bibitem [{\citenamefont {Motohashi}\ \emph {et~al.}(2016)\citenamefont
  {Motohashi}, \citenamefont {Noui}, \citenamefont {Suyama}, \citenamefont
  {Yamaguchi},\ and\ \citenamefont {Langlois}}]{Motohashi:2016ftl}%
  \BibitemOpen
  \bibfield  {author} {\bibinfo {author} {\bibfnamefont {H.}~\bibnamefont
  {Motohashi}}, \bibinfo {author} {\bibfnamefont {K.}~\bibnamefont {Noui}},
  \bibinfo {author} {\bibfnamefont {T.}~\bibnamefont {Suyama}}, \bibinfo
  {author} {\bibfnamefont {M.}~\bibnamefont {Yamaguchi}}, \ and\ \bibinfo
  {author} {\bibfnamefont {D.}~\bibnamefont {Langlois}},\ }\href {\doibase
  10.1088/1475-7516/2016/07/033} {\bibfield  {journal} {\bibinfo  {journal}
  {JCAP}\ }\textbf {\bibinfo {volume} {1607}},\ \bibinfo {pages} {033}
  (\bibinfo {year} {2016})},\ \Eprint {http://arxiv.org/abs/1603.09355}
  {arXiv:1603.09355 [hep-th]} \BibitemShut {NoStop}%
\bibitem [{\citenamefont {Crisostomi}\ \emph {et~al.}(2016)\citenamefont
  {Crisostomi}, \citenamefont {Koyama},\ and\ \citenamefont
  {Tasinato}}]{Crisostomi:2016czh}%
  \BibitemOpen
  \bibfield  {author} {\bibinfo {author} {\bibfnamefont {M.}~\bibnamefont
  {Crisostomi}}, \bibinfo {author} {\bibfnamefont {K.}~\bibnamefont {Koyama}},
  \ and\ \bibinfo {author} {\bibfnamefont {G.}~\bibnamefont {Tasinato}},\
  }\href {\doibase 10.1088/1475-7516/2016/04/044} {\bibfield  {journal}
  {\bibinfo  {journal} {JCAP}\ }\textbf {\bibinfo {volume} {1604}},\ \bibinfo
  {pages} {044} (\bibinfo {year} {2016})},\ \Eprint
  {http://arxiv.org/abs/1602.03119} {arXiv:1602.03119 [hep-th]} \BibitemShut
  {NoStop}%
\bibitem [{\citenamefont {Allys}\ \emph
  {et~al.}(2016{\natexlab{c}})\citenamefont {Allys}, \citenamefont {Peter},\
  and\ \citenamefont {Rodriguez}}]{Allys:2016kbq}%
  \BibitemOpen
  \bibfield  {author} {\bibinfo {author} {\bibfnamefont {E.}~\bibnamefont
  {Allys}}, \bibinfo {author} {\bibfnamefont {P.}~\bibnamefont {Peter}}, \ and\
  \bibinfo {author} {\bibfnamefont {Y.}~\bibnamefont {Rodriguez}},\ }\href
  {\doibase 10.1103/PhysRevD.94.084041} {\bibfield  {journal} {\bibinfo
  {journal} {Phys. Rev.}\ }\textbf {\bibinfo {volume} {D94}},\ \bibinfo {pages}
  {084041} (\bibinfo {year} {2016}{\natexlab{c}})},\ \Eprint
  {http://arxiv.org/abs/1609.05870} {arXiv:1609.05870 [hep-th]} \BibitemShut
  {NoStop}%
\bibitem [{\citenamefont {Jiménez}\ and\ \citenamefont
  {Heisenberg}(2016)}]{Jimenez:2016upj}%
  \BibitemOpen
  \bibfield  {author} {\bibinfo {author} {\bibfnamefont {J.~B.}\ \bibnamefont
  {Jiménez}}\ and\ \bibinfo {author} {\bibfnamefont {L.}~\bibnamefont
  {Heisenberg}},\ }\href@noop {} {\  (\bibinfo {year} {2016})},\ \Eprint
  {http://arxiv.org/abs/1610.08960} {arXiv:1610.08960 [hep-th]} \BibitemShut
  {NoStop}%
\bibitem [{\citenamefont {Golovnev}\ \emph {et~al.}(2008)\citenamefont
  {Golovnev}, \citenamefont {Mukhanov},\ and\ \citenamefont
  {Vanchurin}}]{Golovnev:2008cf}%
  \BibitemOpen
  \bibfield  {author} {\bibinfo {author} {\bibfnamefont {A.}~\bibnamefont
  {Golovnev}}, \bibinfo {author} {\bibfnamefont {V.}~\bibnamefont {Mukhanov}},
  \ and\ \bibinfo {author} {\bibfnamefont {V.}~\bibnamefont {Vanchurin}},\
  }\href {\doibase 10.1088/1475-7516/2008/06/009} {\bibfield  {journal}
  {\bibinfo  {journal} {JCAP}\ }\textbf {\bibinfo {volume} {0806}},\ \bibinfo
  {pages} {009} (\bibinfo {year} {2008})},\ \Eprint
  {http://arxiv.org/abs/0802.2068} {arXiv:0802.2068 [astro-ph]} \BibitemShut
  {NoStop}%
\bibitem [{\citenamefont {Armendariz-Picon}(2004)}]{ArmendarizPicon:2004pm}%
  \BibitemOpen
  \bibfield  {author} {\bibinfo {author} {\bibfnamefont {C.}~\bibnamefont
  {Armendariz-Picon}},\ }\href {\doibase 10.1088/1475-7516/2004/07/007}
  {\bibfield  {journal} {\bibinfo  {journal} {JCAP}\ }\textbf {\bibinfo
  {volume} {0407}},\ \bibinfo {pages} {007} (\bibinfo {year} {2004})},\ \Eprint
  {http://arxiv.org/abs/astro-ph/0405267} {arXiv:astro-ph/0405267 [astro-ph]}
  \BibitemShut {NoStop}%
\bibitem [{\citenamefont {Landau}\ and\ \citenamefont
  {Lifchitz}(1991)}]{Landau:1991wop}%
  \BibitemOpen
  \bibfield  {author} {\bibinfo {author} {\bibfnamefont {L.~D.}\ \bibnamefont
  {Landau}}\ and\ \bibinfo {author} {\bibfnamefont {E.~M.}\ \bibnamefont
  {Lifchitz}},\ }\href@noop {} {\emph {\bibinfo {title} {{Quantum
  Mechanics}}}},\ \bibinfo {series} {Course of Theoretical Physics}, Vol.\
  \bibinfo {volume} {v.3}\ (\bibinfo  {publisher} {Butterworth-Heinemann},\
  \bibinfo {address} {Oxford},\ \bibinfo {year} {1991})\BibitemShut {NoStop}%
\end{thebibliography}
\end{document}